\documentclass[aps,prb,amsmath ,amssymb, twocolumn, superscriptaddress, longbibliography]{revtex4-2}

\newcommand{\ve}[1]{\boldsymbol{#1}}
\renewcommand{\vec}[1]{\boldsymbol{#1}}

\newcommand{\ket}[1]{\left| #1 \right>}
\newcommand{\bra}[1]{\left< #1 \right|}
\newcommand{\braket}[2]{\left< #1 | #2 \right>}

\newcommand{\cfd}{\hat{c}^\dagger}
\newcommand{\f}{\hat{f}}
\newcommand{\fd}{\hat{f}^\dagger}

\renewcommand{\S}{\hat{S}}

\newcommand{\D}[1]{\mathcal{D}\{#1\}}
\newcommand{\hc}{\text{h.c.}}

\usepackage{lipsum}
\usepackage{graphicx,epsfig} 
\usepackage{amsmath}
\usepackage{amsfonts}
\usepackage{cancel}
\usepackage[colorlinks=true, allcolors=blue]{hyperref}
\usepackage{soul}
\usepackage{xcolor}
\usepackage{float}
\usepackage[normalem]{ulem}
\usepackage{orcidlink}

\begin{document}

\title{Unconstrained compact lattice QED\(_{2+1}\) coupled to phonons: Gauss sectors, orthogonal semimetal, and deconfined criticality}

\author{Jo\~ao C. In\'acio \orcidlink{0009-0006-5457-3711}}
\email{joao.carvalho-inacio@uni-wuerzburg.de}
\affiliation{\mbox{Institut f\"ur Theoretische Physik und Astrophysik, Universit\"at W\"urzburg, 97074 W\"urzburg, Germany}}

\author{Fakher F. Assaad \orcidlink{0000-0002-3302-9243}}
\affiliation{\mbox{Institut f\"ur Theoretische Physik und Astrophysik, Universit\"at W\"urzburg, 97074 W\"urzburg, Germany}}
\affiliation{\mbox{W\"urzburg-Dresden Cluster of Excellence ctd.qmat, Germany}}

\date{\today}

\begin{abstract}
  Gauge theoretic descriptions of 2D quantum magnets are defined on a restricted physical Hilbert space by imposing Gauss's law. In this work, we study an unconstrained compact lattice QED\(_{2+1}\) coupled to bond phonons, in which local Gauss operators are conserved, but their eigenvalues are not \textit{a priori} fixed. Using exact auxiliary-field quantum Monte Carlo simulations at \(N_f = 2\), we investigate how the physical Gauss sector becomes energetically favourable, and how this process is connected to confinement and symmetry breaking. We identify an orthogonal semimetallic (OSM) phase, in which fermions striped off their gauge charge and form a Dirac liquid.  This phase mixes various Gauss sectors, and  becomes unstable when Gauss's law is explicitly enforced. Upon the imposition of Gauss's law, compactness allows for monopole excitations carrying antiferromagnetic (AFM) and valence-bond-solid (VBS) quantum numbers. Gauge field fluctuations and spinon-phonon coupling tune between the competition of such monopoles and generate a phase diagram with OSM, AFM and VBS orders. The transitions out of the OSM phase coincide with the dynamical generation of Gauss's law, while the competition between AFM and VBS charged monopoles produces a transition consistent with deconfined quantum criticality. Thus our results establish that upon the projection to the physical Hilbert space, deconfinement in compact lattice QED\(_{2+1}\) is always unstable towards either  AFM or VBS order.
\end{abstract}

\maketitle

\section{Introduction}

Gauge theories provide a powerful framework to describe emergent degrees of freedom in condensed matter systems. As such, they are central to our understanding of many correlated systems and phenomena, for example, heavy fermionic systems \cite{saremi_quantum_2007,raczkowski_breakdown_2022,pan_quantum_2025}, superconductivity \cite{baskaran_gauge_1988,zhang_unified_1997}, spin systems \cite{affleck_large-n_1988,wen_quantum_2002,hermele_algebraic_2005,senthil_competing_2006} and deconfined quantum criticality \cite{senthil_quantum_2004,senthil_deconfined_2004,wang_deconfined_2017}. In particular, unordered spin states provide an interesting setting in which gauge structures emerge as low-energy degrees of freedom. A canonical example is the Kitaev model \cite{kitaev_anyons_2006}, where \(S=1/2\) spins are explicitly fractionalised into Majorana fermions hopping in the background of a \(\mathbb{Z}_2\) gauge field. More generally, in parton constructions of quantum magnets, the physical spin is represented in an enlarged fermionic Hilbert space and the accompanying redundancy produces a local gauge symmetry. Recovering the spin model requires a local constraint, or equivalently Gauss's law. Imposing Gauss's law projects the enlarged Hilbert space to the physical spin one. A gauge invariant Hamiltonian only guarantees that the local Gauss operators are conserved; it does not require the calculation to remain in a single sector. If  Gauss sectors are degenerate, unphysical sectors contribute to low-energy physics. This raises two related questions: whether the physical sector is selected dynamically, and what phases can occur when the gaps between different sectors collapse.

\begin{figure}[b]
  \centering
  \includegraphics[width=.45\textwidth]{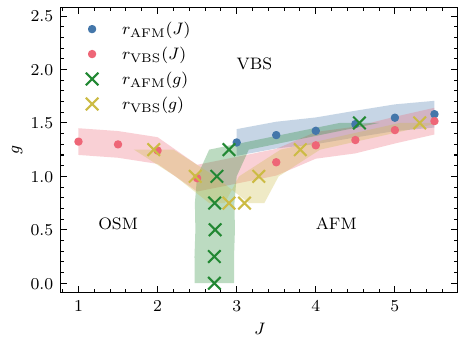}
  \caption{Ground state phase diagram of the unconstrained Hamiltonian in Eq.~\eqref{eq:hamiltonian} as a function of the electric field strength \(J\) and spinon-phonon coupling \(g\), for \(\omega_0 = 2\) and \(K = 1\). The diagram contains OSM, AFM, and VBS phases. The positions of the phase boundaries are approximately determined from the crossings of the AFM and VBS correlation ratio for finite-size data. The shaded regions are an estimate of the error, i.e. half of the grid width.}
  \label{fig:phase_diagram}
\end{figure}

The square lattice U(1) parton construction of the Heisenberg model leads at low energies to QED\(_{2+1}\), in which massless Dirac spinons are coupled to an emergent U(1) gauge field \cite{hermele_stability_2004,hermele_algebraic_2005,song_unifying_2019}. In the non-compact theory, the U(1) Dirac spin liquid (DSL) has an enlarged SU(4) flavour symmetry. Compactness, however, permits instanton events that insert \(2\pi\) gauge flux, described by monopole operators. On the square lattice, the monopole multiplet contains components with antiferromagnetic (AFM) and valence-bond-solid (VBS) quantum numbers. Monopole proliferation confines the spinons and selects either ordered state \cite{alicea_monopole_2008,song_unifying_2019,song_spinon_2020}. 

The competition between these monopoles is especially significant at a direct AFM-VBS transition. As the two phases break unrelated symmetries, a continuous transition between them lies beyond the conventional Landau-Ginzburg-Wilson (LGW) paradigm and is described by a deconfined quantum critical point (DQCP) \cite{senthil_quantum_2004,senthil_deconfined_2004}. Theories for DQCP are characterised either by fractionalised bosonic spinons coupled to a non-compact U(1) gauge field, or by a fermionic descriptions where three AFM and two VBS components into a five-component vector with an emergent SO(5) symmetry \cite{senthil_competing_2006,tanaka_many-body_2005,wang_deconfined_2017,nahum_emergent_2015,SatoT22}. This relation also connects U(1) and SU(2) parton formulations and provides a controlled setting in which to examine the link between gauge theories and deconfined criticality \cite{song_spinon_2020,takahashi_so5_2024}.

Here we study an unconstrained compact U(1) lattice gauge theory coupled to bond phonons on the square lattice using exact auxiliary-field quantum Monte Carlo simulations at \(N_f = 2\) fermion flavours. In contrast to earlier interpretations which argue for a stable DSL phase \cite{xu_monte_2019,feng_scalable_2026}, we find that the gaps between Gauss sectors collapse in this parameter regime. We identify this phase as an orthogonal semimetal (OSM): gauge invariant spin and dimer correlations retain signatures of free Dirac fermions, but the physical Gauss sector is not energetically isolated. Explicitly imposing the constraint removes this phase and yields AFM order, showing that upon the projection to the physical sector the DSL is not stable. The resulting phase diagram, shown in Fig. \ref{fig:phase_diagram}, contains OSM, AFM, and VBS phases. At both OSM-AFM and OSM-VBS phase boundaries Gauss's law becomes energetically imposed. Enhanced VBS (AFM) fluctuations at the OSM-AFM (OSM-VBS) transition, support the claim of deconfinement at criticality. We also find evidence for a continuous AFM-VBS transition with enlarged SO(5) symmetry, i.e. DQCP. 

The paper is organised as follows. In Sec. \ref{sec:model}, we define the model and discuss its symmetries. In Sec. \ref{sec:limits} we analyse the limits of the model, in particular we discuss the OSM and its stability to gauge fluctuations. Then in Sec. \ref{sec:results} we present our results. We start by discussing the whole phase diagram and then focus on the physical properties of the OSM  phase, confinement transitions and AFM-VBS transition. Finally, in Sec. \ref{sec:conclusion} we conclude our study.

\section{Model, symmetries and method}
\label{sec:model}

\subsection{Compact lattice QED\(_{2+1}\)}

We are interested in studying a compact U(1) lattice gauge theory coupled to phonons. Our model is described by the Hamiltonian:
\begin{widetext}
  \begin{equation} \label{eq:hamiltonian}
    \hat{H} = \sum_{i,\delta,\sigma} (-t + g \hat{X}_{i,\delta}) \left(\f^\dagger_{i,\sigma} \hat{U}_{i,\delta} \f_{i+\delta,\sigma} + \hc\right) + J \sum_{i,\delta} \frac{\hat{E}^2_{i,\delta}}{4} + K \sum_\square \cos(\text{curl} \hat{A}_\square) + \sum_{i,\delta} \left(\frac{\hat{P}^2_{i,\delta}}{2m} + \frac{k}{2} \hat{X}_{i,\delta}^2\right),
  \end{equation}
\end{widetext}
where \(\hat{U}_{i,\delta} = e^{i\hat{A}_{i,\delta}}\), \([\hat{A}_{i,\delta}, \hat{E}_{i^\prime,\delta^\prime}] = [\hat{X}_{i,\delta}, \hat{P}_{i^\prime,\delta^\prime}] = i \delta_{i,i^\prime} \delta_{\delta, \delta^\prime}\) and \([\hat{E}_{i,\delta}, \hat{U}_{i',\delta'}] = \delta_{i,i'} \delta_{\delta,\delta'} \hat{U}_{i,\delta}\). Here \(\hat{A}_{i,\delta}\) and \(\hat{E}_{i,\delta}\) are the gauge and electric field variables, \(\hat{X}_{i,\delta}\) and \(\hat{P}_{i,\delta}\) are the phonon position and momentum operators, and \(\hat{U}_{i,\delta}\) is a link operator. Both of these live on a bond \(b = (i,\delta)\) which connects sites \(i\) to \(i+\delta\) (\(\delta = x,y\)) on a \(N = L \times L\) square lattice. The Hilbert space of \(\hat{A}_{i,\delta}\) is \(L^2([0, 2\pi])\), defining a compact U(1) gauge field. The operators \(\{\f_{i,\sigma}, \fd_{j,\sigma'}\} = \delta_{i,j}\delta_{\sigma,\sigma'}\) are the fermionic spinon operators and \(\sigma =\ \uparrow,\downarrow\) is the flavour/spin index. \(J\) controls the dynamics of the gauge fields and \(K > 0\) is the flux term which favours a \(\pi\)-flux on each plaquette \(\square\). The magnetic flux is defined as \(\text{curl} \hat{A}_\square = \sum_{b\in\square} \hat{A}_b\), where the summation is oriented (anti)clockwise. The spinon-phonon coupling is given by \(\lambda = g^2/2k\) and the phonon frequency is \(\omega_0 = \sqrt{k/m}\). Without loss of generality, we set \(k = 2\). For \(g = 0\), this model has been extensively studied as a function of \(J\) and fermion flavours in Refs.~\cite{xu_monte_2019,janssen_confinement_2020,wang_dynamics_2019,feng_scalable_2026,chen_emergent_2026}.

Our model is justified from the point of view of a parton mean-field (MF) description of a Heisenberg model on the square lattice. Each spin is represented by two fermionic spinons \(\vec{\hat{S}}_i = \frac{1}{2} \fd_{i,\sigma} \vec{\sigma}_{\sigma\sigma'} \f_{i,\sigma'}\), subjected to a local constraint \(\hat{Q}_i = \sum_\sigma \fd_{i,\sigma}\f_{i,\sigma} - 1 = 0\). This redundancy implies an emergent U(1) gauge structure which becomes evident in the coherent state path integral formalism \cite{wen_quantum_2002,hermele_algebraic_2005,song_unifying_2019}. Note that the parton description is also valid when considering a coupling to phonons as these do not break local U(1) symmetry. In the continuum fermionic spinons are minimally coupled to a non-compact U(1) gauge field in 2+1 dimensions. The resulting theory is a version of QED\(_{2+1}\) with four fermion flavours \cite{song_unifying_2019}. So, the original SU(2) flavour symmetry is enlarged to SU(4) due to extra degrees of freedom. An important feature of the low-energy theory is the appearance of an additional conserved quantity associated with the magnetic flux \cite{borokhov_topological_2003,hermele_algebraic_2005,song_unifying_2019}, giving rise to a topological U(1) symmetry. It has no counterpart in the original microscopic model as it emerges due to the compactness of the gauge field. This symmetry is associated with instanton events which modify the gauge field in space-time by \(\pm2\pi\), also known as monopoles. Monopoles transform nontrivially under lattice symmetries \cite{alicea_monopole_2008,song_unifying_2019}, and their condensation \cite{polyakov_quark_1977} dynamically generates a Dirac mass whose symmetry properties are fixed by those transformations, confining the gauge theory.

\subsection{Symmetries and Gauss's law}

\begin{figure}[t]
  \centering
  \includegraphics[width=.45\textwidth]{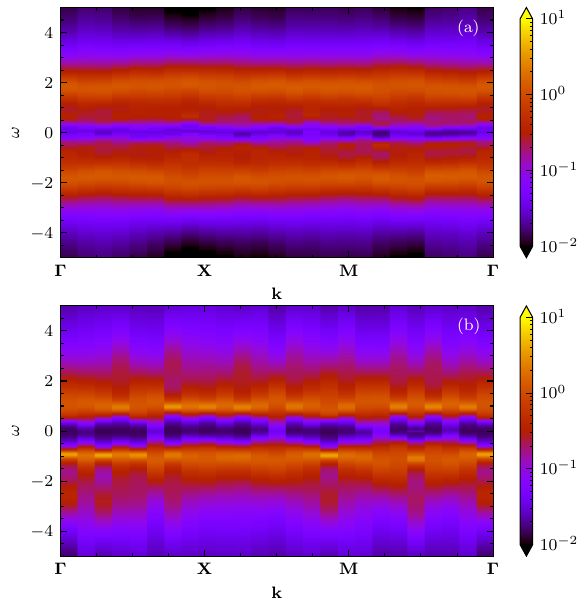}
  \caption{Representative single particle spectral function \(A(\vec{k},\omega)\) at (a) \(J = 1\) and (b) \(J = 5\), both at \(g = 0\), obtained by SAC. The colour scale is logarithmic. Here \(\beta = L = 16\).}
  \label{fig:greens_function_spectral}
\end{figure}

The model defined in Eq.~\eqref{eq:hamiltonian} leads to a number of symmetries. It has time-reversal, spin SU(2) and all of the lattice translation and C\(_4\) rotation symmetries. The model also has particle-hole and local U(1) symmetries which lead to interesting consequences. 

Under a particle-hole transformation,  
\begin{equation}
  \hat{\mathcal{P}} z \fd_{i,\sigma} \hat{\mathcal{P}}^{-1} = z^* (-1)^i \f_{i,\sigma},
\end{equation}
where \((-1)^i\) is \(1\) (\(-1\)) if \(i\) is in sublattice \(A\) (\(B\)) and \(z \in \mathbb{C}\). Under this transformation, the hopping term transforms as  
\begin{equation}
  \hat{\mathcal{P}} \fd_{i,\sigma} \hat{U}_{i,\delta} \f_{i+\delta, \sigma} \hat{\mathcal{P}}^{-1} = - \f_{i,\sigma} \hat{\mathcal{P}} \hat{U}_{i,\delta} \hat{\mathcal{P}}^{-1} \fd_{i+\delta,\sigma}.
\end{equation}
Then \(\hat{\mathcal{P}} \hat{U}_{i,\delta} \hat{\mathcal{P}}^{-1} = \hat{U}_{i,\delta}^\dagger\). As \(\hat{E}_{i,\delta} = -i \partial_{\hat{A}_{i,\delta}}\), under a particle-hole transformation, the electric field operator transforms as \(\hat{\mathcal{P}} \hat{E}_{i,\delta} \hat{\mathcal{P}}^{-1} = - \hat{E}_{i,\delta}\). This symmetry has the consequence that \(\langle \hat{E}_{i,\delta} \rangle = 0\).  

Next we investigate the local U(1) symmetry and its consequences. The generator of a local U(1) transformation is given by 
\begin{equation}
  \hat{T}_{Q} = e^{i \sum_i \theta_i \hat{Q}_i},
\end{equation}
with \(\hat{Q}_i = - \sum_{\delta} (\hat{E}_{i,\delta} - \hat{E}_{i, -\delta}) + \sum_{\sigma} \fd_{i,\sigma} \f_{i,\sigma} - 1\). One can show that 
\begin{gather}
  \hat{T}_{Q} e^{i \hat{A}_{i,\delta}} \hat{T}_{Q}^{-1} = e^{i \hat{A}_{i,\delta}} e^{i (\theta_{i+\delta} - \theta_i)}, \\
  \hat{T}_{Q} \f_{i,\sigma} \hat{T}_{Q}^{-1} = e^{-i\theta_{i}} \f_{i,\sigma}.
\end{gather}
Then the Hamiltonian is invariant under a local U(1) gauge transformation, i.e. \([\hat{H}, \hat{Q}_i] = 0 \). The quantity \(\hat{Q}_i\) is a restatement of Gauss's law and it is a conserved quantity and the eigenvalues define (gauge) charge sectors, here named Gauss sectors. This implies that the eigenstates of the Hamiltonian \(\left| \Psi \right>\) can then be written as 
\begin{equation} \label{eq:eigenstates_gauss_sectors}
  \hat{Q}_i \left| \Psi(Q_1, \hdots, Q_N) \right> = Q_i \left| \Psi(Q_1, \hdots, Q_N) \right>.
\end{equation}
The physical Hilbert space of the spin wave functions is defined for \(Q_i = 0\) for all \(i\) \cite{wen_quantum_2002}. In this sector, the divergence of the electric field around site \(i\) is related its local charge. Satisfying Gauss's law corresponds to locally imposing the constraint \(Q_i = 0\) or equivalently projecting to the physical Hilbert space. In our model we do not impose Gauss's law, such that our Hamiltonian describes an \textit{unconstrained} gauge theory. The operators \(\hat{Q}_i\) are conserved quantities but their eigenvalues are not fixed. This implies that any gauge dependent observable is local in space but not in imaginary-time, i.e. \(\langle \fd_{i,\sigma}(\tau) \f_{j,\sigma} (0) \rangle = \delta_{i,j} g(\tau)\), where \(g(\tau)\) is some non-trivial time dependence. So, the \(f\)-fermion can not propagate in real space. Fig.~\ref{fig:greens_function_spectral} shows that this quantity is \(\vec{k}\)-independent. However, quantities such as 
\begin{equation}
  \langle \f_{i,\sigma} \hat{U}_{i,i_1} \hdots \hat{U}_{i_n, j} \fd_{j,\sigma} \rangle
\end{equation}
have a non-zero value. The commutator \([ \hat{Q}_i, \hat{f}^{\dagger}_{j,\sigma} ] = \delta_{i,j} \hat{f}^{\dagger}_{j,\sigma}\) picks up the gap between the ground state Gauss law sector, say $Q_1, \cdots Q_N$, and $Q_1, \cdots Q_j+1, \cdots Q_N$. If the gap is finite, then the Gauss law is imposed energetically. For energies below the gap to the first excited gauge sector, our theory maps onto a genuine \textit{constrained} gauge theory: applying an operator on the ground state that alters the Gauss law will produce an excited state, with exponentially small Boltzmann weight. 

Since the $\hat{Q}_i$ are conserved quantities, the energetics of the Gauss sectors will be determined by a classical Hamiltonian dictated by symmetry:
\begin{equation}
  \hat{H}_Q  = \sum_{i, j} J_Q(i - j) \hat{Q}_i \hat{Q}_j  +  \cdots
\end{equation}
Here we have  assumed translation  symmetry and since $ \mathcal{\hat{P}} \hat{Q}_i \mathcal{\hat{P}}^{-1} = - \hat{Q}_i$, only even powers of $\hat{Q}_i$ are allowed in the effective Hamiltonian.

\subsection{Method}

To simulate our model, Eq.~\eqref{eq:hamiltonian}, we use a modified version of the Algorithms for Lattice Fermions \cite{assaad_alf_2025} implementation of the auxiliary-field quantum Monte Carlo (AF-QMC) method. The details of this implementation can be found in Appendix \ref{app:af_qmc}. It is worth noting that we use a local updating scheme based on the Metropolis-adjusted Langevin algorithm (MALA) \cite{roberts_exponential_1996} to simultaneously update the gauge and phonon fields in our simulation. For our simulations we use an asymmetric Trotter decomposition with a Trotter step of \(\Delta\tau = 0.1\) and we set the Langevin time step \(\delta t_{\text{g}/\text{p}}\) for the gauge and phonon fields such that the acceptance ratio is 70-80\% to reduce autocorrelation times.

\section{Limits of the model}
\label{sec:limits}

Here we present the different limits of the model as a function of electric field fluctuations \(J\) and phonon frequency \(\omega_0\). In particular we will discuss the limit of small gauge fluctuations, \(J \to 0\), in depth.

\subsection{\(J = 0\) case}

At \(J = 0\), the fluctuations of the gauge fields are absent, and thus they become classical variables. In this limit, the Hamiltonian, without phonons, is given by 
\begin{equation}
  \hat{H} = - t \sum_{i,\delta,\sigma} \left(\f^\dagger_{i,\sigma} e^{iA_{i,\delta}} \f_{i+\delta,\sigma} + \hc\right) + K \sum_\square \cos(\text{curl} A_\square).
\end{equation}
For \(K > 0\), \(\text{curl} A_\square = \pm \pi\) minimised the cosine term, and a \(\pi\)-flux per unit cell is dynamically generated. Owing to Lieb's theorem 
\cite{lieb_flux_1994}, this flux configuration is energetically favoured even at $K=0$. Thus the resulting Hamiltonian is given by 
\begin{equation}
  \hat{H} = - t \sum_{i,\delta,\sigma} e_{i,\delta} \left(\f^\dagger_{i,\sigma} \f_{i+\delta,\sigma} + \hc\right),
\end{equation}
where \(e_{i,\delta}\) accounts for the \(\pi\)-flux. This phase is the mean-field description of the DSL \cite{wen_quantum_2002}, where gauge fluctuations are not present. The constraint here reduces to \(\hat{Q}_i = \sum_{\sigma} \fd_{i,\sigma} \f_{i,\sigma} - 1\). This point is singular in the sense that a local symmetry, the local U(1) symmetry, is broken. As a consequence, the \(f\)-fermions acquire a dispersion.

\subsection{\(J \to 0\) limit}

As \(J \to 0\), the fluctuations of the gauge field become progressively slower than the time-scale of the fermions, and Gauss's law is not energetically imposed. Thus the fermions decouple from the gauge fields, and both quantities retain gapless excitations. Importantly, the local U(1) symmetry is not broken in this state. This is corresponds to the OSM phase. We now study the properties and stability of this phase for vanishing spinon-phonon coupling \(g = 0\). 

It turns out that at \(J \to 0\) the various Gauss law sectors are degenerate. Let
\begin{gather}
  \hat{H} \ket{\Psi(Q_1, \cdots Q_N) } = E \ket{\Psi(Q_1, \cdots Q_N) },  \\
  \hat{Q}_{i} \ket{\Psi(Q_1, \cdots Q_N) } = Q_i \ket{\Psi(Q_1, \cdots Q_N) }.
\end{gather}
Consider a  gauge field string with open ends: 
\begin{equation}
  \hat{W}_{i,j}  = \hat{U}_{i,i_1} \hat{U}_{i_1,i_2} \cdots \hat{U}_{i_n,j}.
\end{equation} 
At \(J \to 0\), the gauge field string  $\hat{W}_{i,j}$  commutes with the Hamiltonian. Thus \(\hat{W}_{i,j} \ket{\Psi(Q_1, \cdots Q_N) }\) is an eigenstate of the Hamiltonian with the same energy \(E\) but with \(Q_i \to Q_i+1\) and \(Q_j \rightarrow Q_j-1\). So excitations between Gauss sectors require no finite energy, i.e. Gauss's law is not energetically imposed at \(J \to 0\). Since gauge field strings with open ends cost no energy at \(J \to 0\) and leave \( \hat{Q}_{\text{tot}} = \sum_i \hat{Q}_i \) unchanged, the energetics of the Gauss sectors is governed by a term of the form
\begin{equation} \label{eq:HQ_Higgs}
  \hat{H}_{Q} = \frac{\lambda_{Q}}{N}  \left( \sum_i \hat{Q}_i \right)^2  + \cdots 
\end{equation}
Here \(N\) is the total number of sites, and this normalisation ensures that the energy remains extensive. Aside from the degeneracy among Gauss sectors sharing the same total charge \( \hat{Q}_{\text{tot}} \), the energy difference between Gauss sectors  with \( \hat{Q}_{\text{tot}} \) differing by unity scales as \( \lambda_{Q}/N \). As a consequence of this vanishing gap in the thermodynamic limit, we expect the susceptibility associated with the fluctuations of \( \hat{Q}_{\text{tot}} \) to show no activated behaviour at low temperatures. This is a numerical hallmark of this state.

To understand the nature of the ground state at \(J \to 0\), it is convenient to adopt  a  parton construction and write the \(f\)-fermions in terms of slave bosons and \(c\)-fermions:
\begin{equation}  \label{eq:slave_boson}
  \f^{\dagger}_{i,\sigma} = \cfd_{i,\sigma} \hat{b}_i
\end{equation}
We adopt the state identification 
\begin{align}
  \hat{f}^{\dagger}_{\sigma} \ket{0} \equiv \ket{\sigma}   &\leftrightarrow \ket{1,\sigma}   \equiv  \hat{b}^{\dagger} \hat{c}^{\dagger}_{\sigma} \ket{0} \\
  \ket{0} &\leftrightarrow \ket{2,0} \equiv (\hat{b}^{\dagger})^2 \ket{0} \\
  \hat{f}^{\dagger}_{\uparrow} \hat{f}^{\dagger}_{\downarrow} \ket{0}  \equiv \ket{\uparrow\downarrow} &\leftrightarrow \ket{0, \uparrow\downarrow} \equiv \hat{c}^{\dagger}_{\uparrow} \hat{c}^{\dagger}_{\downarrow}  \ket{0}
\end{align}
that naturally  leads  to the constraint: 
\begin{equation} \label{eq:parton_constraint}
  \hat{b}^{\dagger}_i \hat{b}_i + \sum_{\sigma} \hat{c}^{\dagger}_{i,\sigma} \hat{c}_{i,\sigma} = 2.
\end{equation}
With the above, 
\begin{equation}
  \hat{Q}_i = - \sum_{\delta} (\hat{E}_{i,\delta} - \hat{E}_{i, -\delta}) +  1 -  \hat{b}^{\dagger}_i \hat{b}_i.
\end{equation}
Since \(\hat{Q}_i\) is the generator of the local U(1) gauge transformation, the bosons carry the gauge charge while the \(c\)-fermions are gauge neutral. As a consequence the \(c\)-fermions, unlike the \(f\)-fermions, are free to propagate in space. We will follow the terminology of Refs.~\cite{Nandkishore12,Hohenadler18,Hohenadler19} and refer to the \(f\)-fermions as orthogonal fermions. 

Within this parton construction, the original Hamiltonian, Eq.~\eqref{eq:hamiltonian}, is written as
\begin{multline} \label{eq:hamiltonian_sb}
  \hat{H} = -t \sum_{i,\delta,\sigma} \left(\hat{c}^\dagger_{i,\sigma} \hat{b}_i e^{i\hat{A}_{i,\delta}} \hat{b}^{\dagger}_{i+\delta} \hat{c}_{i+\delta,\sigma} + \hc \right) + \\
  J \sum_{i,\delta} \frac{\hat{E}^2_{i,\delta}}{4} + K \sum_\square \cos(\text{curl} \hat{A}_\square). 
\end{multline}
In the limit \(J \to 0\), the gauge field progressively freezes out, i.e. quantum fluctuations of the gauge field become increasingly slow. The quantity \(\hat{b}_i e^{i\hat{A}_{i,\delta}} \hat{b}^{\dagger}_{i+\delta}\) carries no gauge charge such that the fluctuations of the bosons are tied to those of the gauge fields. Hence in this limit, we can replace the operator \(\hat{b}_i e^{i\hat{A}_{i,\delta}} \hat{b}^{\dagger}_{i+\delta}\) by a number:
\begin{equation}
  \hat{b}_i e^{i\hat{A}_{i,\delta}} \hat{b}^{\dagger}_{i+\delta} \to \chi_{i,\delta} 
\end{equation}
Here, $\chi_{i,\delta}$ is a complex number, the phase of which is set by  Lieb's theorem \cite{lieb_flux_1994} so as to generate a $\pi$-flux per plaquette. We will set $|\chi_{i,\delta}| = 1 $ as obtained from a parton mean-field treatment, in which bosons condense. Hence, $\chi_{i,\delta}  = e_{i,\delta}$  as at $J= 0$,  and   the Hamiltonian reads:
\begin{equation} \label{eq:hamiltonian_J0}
  \hat{H} = -t \sum_{i,\delta,\sigma} \left(\hat{c}^\dagger_{i,\sigma}  e_{i,\delta} \hat{c}_{i+\delta,\sigma} + \hc\right).
\end{equation}
In the above we have omitted the flux term since it merely accounts for a constant energy shift of $-K N$. The Hamiltonian in Eq.~\eqref{eq:hamiltonian_J0} describes free \(c\)-Dirac fermions. We note that in our AF-QMC simulations we do not have access to the \(c\)-fermions but only to the \(f\)-fermions. However, 
\begin{equation} 
  \label{eq:spin_operator_c}
  \ve{\S}_{i} = \frac{1}{2} \hat{f}^{\dagger}_{i,\sigma} \ve{\sigma}_{\sigma,\sigma'} \hat{f}_{i,\sigma'}  \simeq  \frac{1}{2} \hat{c}^{\dagger}_{i,\sigma}  \ve{\sigma}_{\sigma,\sigma'} \hat{c}_{i,\sigma'}
\end{equation} 
where the proportionality  factor is  given by  \(\langle \hat{b}_i \hat{b}^{\dagger}_i \rangle \). Hence, gauge invariant quantities, such as spin-spin correlation functions, share the same properties. 

The key point in this derivation, is that we have broken translation symmetry due to the emergent $\pi$-flux pattern. In particular, 
\begin{equation}
  \hat{T}_{x} \hat{T}_{y} = - \hat{T}_{y} \hat{T}_{x}
\end{equation} 
such that the unit cell has doubled and the Dirac nature of the \(c\)-fermions does not violate Luttinger's volume \cite{oshikawa_topological_2000}. This however does not imply that the gauge fields are frozen, i.e. the slow gauge fluctuations are compensated by the fluctuations of the \(b\)-bosons such that the \(\pi\)-flux pattern remains. We will refer to this state as an OSM \cite{gazit_fermi_2020,nandkishore_orthogonal_2012}. This is a distinct state from the DSL and the \(\pi\)-flux states. In the former, the full translation symmetry is not broken and Gauss's law is energetically imposed, and in the latter, the local U(1) symmetry is absent. An important question is whether the OSM phase is stable to small values of \(J\).

\subsection{Stability of the OSM phase}

For simplicity, we will set \(K = 0\) and ask whether, perturbatively, the OSM phase is stable to small fluctuations of the gauge field, i.e. small-\(J\) regime. To do so, we will start with the slave boson representation introduced in the previous section, and account for  small fluctuations of the gauge-invariant \(e_{i,\delta}\) link variable. Our starting point is the Hamiltonian: 
\begin{equation}
  \hat{H} = -t \sum_{i,\delta}  \left(\hat{c}^\dagger_{i} \hat{b}_i e^{i\hat{A}_{i,\delta}} \hat{b}^{\dagger}_{i+\delta} \hat{c}_{i+\delta} + \hc \right) + \frac{J}{4} \sum_{i,\delta} \hat{E}^2_{i,\delta}.
\end{equation} 
In the above $\hat{c}^\dagger_{i} =  ( \hat{c}^\dagger_{i,\uparrow}, \hat{c}^\dagger_{i,\downarrow} )$ is a two-component spinor. We will work in a coherent state path integral formulation and obtain the following action:
\begin{multline}
  S = S_0 + \int_{0}^{\beta} d\tau \sum_{i,\delta} \Big[ -t (c^\dagger_{i} b_i e^{i a_{i,\delta}} b^{\dagger}_{i+\delta} c_{i+\delta} + \hc ) \\ 
  + \frac{1}{J} (\partial_\tau a_{i,\delta})^2 \Big]
\end{multline}
where 
\begin{equation}
S_0 = \int_{0}^{\beta} d \tau \sum_{i} \Big[  c^\dagger_{i} \partial_\tau c_i +  b^\dagger_{i}\partial_\tau b_i   + i \lambda_i ( c^\dagger_i c_i + b^\dagger_i b_i - 2 ) \Big]
\end{equation}
accounts for the slave boson constraint as well as the fermionic and bosonic Berry phases.  The path integral formulation allows us to separate time scales. With the Fourier transform: 
\begin{equation}
  a_{i,\delta}(\tau) = \frac{1}{\sqrt{\beta}} \sum_{i \Omega_m} a_{i,\delta}(i\Omega_m)  e^{i \Omega_m \tau}
\end{equation}
we define slow and fast components of the gauge field 
\begin{gather}
  a_{i,\delta}^{<}(\tau) = \frac{1}{\sqrt{\beta}} \sum_{ |i \Omega_m| < \Lambda} a_{i,\delta}(i\Omega_m)  e^{i \Omega_m \tau}, \\
  a_{i,\delta}^{>}(\tau) = \frac{1}{\sqrt{\beta}} \sum_{|i \Omega_m| > \Lambda} a_{i,\delta}(i\Omega_m)  e^{i \Omega_m \tau},
\end{gather}
respectively. Here \(\Lambda\) is the energy cut-off which separates the slow and fast modes of the gauge field. The slow mode accounts for the OSM phase, i.e. dynamical generation of the \(\pi\)-flux, such that we make the approximation: 
\begin{equation}
  c^\dagger_{i}(\tau) b_i(\tau) e^{i a^{<}_{i,\delta}(\tau)} b^{\dagger}_{i+\delta}(\tau) c_{i+\delta}(\tau) = c^\dagger_{i}(\tau) e_{i,\delta} c_{i+\delta}(\tau).
\end{equation}
For small values of \(J\), \(a_{i,\delta}^{>}(\tau)\) will be \textit{small}, such that we can expand the exponential to obtain the action:
\begin{multline}
  S_{\text{eff}} = \underbrace{\int_{0}^{\beta} d\tau \Big[ \sum_{i} c^\dagger_{i} \partial_\tau c_i  - t \sum_{i,\delta} (c^\dagger_{i} e_{i,\delta} c_{i+\delta} + \hc )  \Big] }_{\equiv S^<} \\
  + \int_{0}^{\beta} d\tau  \sum_{i,\delta} \left[  a^{>}_{i,\delta} j_{i,\delta}  + \frac{1}{J} (\partial_\tau a^>_{i,\delta})^2 \right].
\end{multline} 
Here the paramagnetic current reads \(\hat{j}_{i,\delta} = -i t (\hat{c}^\dagger_{i} e_{i,\delta} \hat{c}_{i+\delta} - \hc)\). In the above we impose the constraint only on average, and neglect the fluctuations of the bosonic fields. We can now integrate out the fast modes to obtain the effective action:
\begin{equation}
  S_{\text{eff}} = S^< - \underbrace{\frac{J}{4} \sum_{i,\delta} \int d\tau d\tau' j_{i,\delta}(\tau) P(\tau - \tau') J_{i,\delta}(\tau')}_{\equiv S^>}
\end{equation}
where 
\begin{equation}
  P(\tau) = \frac{1}{\beta} \sum_{|i \Omega_m| > \Lambda} \frac{e^{i \Omega_m \tau}}{\Omega_m^2}.
\end{equation}
We can now ask whether the fast modes are relevant at the  Dirac fixed point described by \( S^< \). At this fixed point, the scaling dimension of the fermions is given by \(\Delta_c = 1\) in two dimensions. Hence, under a scale transformation \(\tau = \tilde{\tau} b\) with \(b > 1\), 
\begin{equation}
  S^> \to  b^{2 - 4 \Delta_c + 1} S^>. 
\end{equation}
The  above stems from the fact that \(P(b\tilde{\tau}) = b P(\tilde{\tau})\) for \(\Lambda \to 0\) at \(T = 0\). The above  suggests that the OSM phase is stable to small gauge fluctuations.

\subsection{\(J \to \infty\) limit}

When \(J \to \infty\), our model is maps on to the Heisenberg model \cite{xu_monte_2019}. At \(J = \infty\), cost of creating quantum of electric field becomes infinitely large. The fermions become completely localised. For any finite \(J \to \infty\), electric field excitations are allowed and the fermions become less localised. In this limit we can do an expansion in powers of \(t^2/J\). At \(J = \infty\), the ground state has zero quantum of electric field and \([\hat{E}_{i,\delta}, \hat{U}_{i,\delta}] = \hat{U}_{i,\delta}\), so that a hopping process creates a quanta of electric field. Then the lowest order process at \(J \to \infty\) is two hopping processes where a virtual quantum of electric field is created. In this limit the effective Hamiltonian is written as 
\begin{equation}
  \hat{H}_{\text{eff}} = - \frac{t^2}{J} \sum_{i,\delta} \left( \hat{D}^\dagger_{i,i+\delta} \hat{D}_{i,i+\delta} + \hat{D}_{i,i+\delta} \hat{D}^\dagger_{i,i+\delta} \right),
\end{equation}
where \(\hat{D}_{i,i+\delta} = \sum_\sigma \fd_{i, \sigma} \f_{i+\delta, \sigma}\). So for \(J\to\infty\) the low energy processes of our lattice gauge theory model are encoded in the antiferromagnetic Heisenberg model with exchange constant \(J_{\text{Heisenberg}} \sim t^2 / J\). Note that in this limit, Gauss's law is dynamically imposed as the only allowed processes are virtual hoppings, where a quanta of electric field is created, in the background of zero electric field.

In particular, in the large $J$ limit, the energetics of the Gauss sectors is expected to follow from the Hamiltonian:
\begin{equation}
  \hat{H}_{Q} = J_{Q} \sum_{i}  \hat{Q}_{i} ^2  + \cdots
\end{equation}
This classical Hamiltonian has a unique ground state corresponding to \(\hat{Q}_i = 0\) for all \(i\), which enforces Gauss's law dynamically. Excitations away from this ground state come at an energy cost set by $J_Q$. Hence, in this state, the susceptibility associated with fluctuations of $\hat{Q}_{\text{tot}} $ will show activated behaviour in the low-temperature limit. This is a hallmark of a dynamically imposed Gauss law.

\subsection{Adiabatic and antiadiabatic phonon limits}

Now we study the limits of the Hamiltonian Eq.~\eqref{eq:hamiltonian} at finite spinon-phonon coupling \(g > 0\) as a function of \(\omega_0\). 

In the adiabatic limit, \(\omega_0 = 0\), the phonons become classical variables as quantum fluctuations are frozen. Thus we can write the phonons in the position basis \(\hat{X}_{i,\delta} \ket{x} = x_{i,\delta} \ket{x}\). At \(J = 0\), however, the gauge fields dynamically generate a \(\pi\)-flux pattern and the mean-field Hamiltonian is given by 
\begin{equation}
  \hat{H} = \sum_{i,\delta,\sigma} (-t + g x_{i,\delta}) e_{i,\delta} (\fd_{i,\sigma} \f_{i+\delta,\sigma} + \hc) + \frac{k}{2} \sum_{i,\delta} x_{i,\delta}^2. 
\end{equation}
Thus, as a function of \(g\), the phonons can condense and generate a \((\pi, 0)\) VBS order \cite{ryu_masses_2009,herbut_so8_2023}. This transition belongs to the large-$N$ Gross-Neveu-XY universality class.

On the other hand, the antiadiabatic limit, \(\omega_0 \to \infty\), the spinon-phonon interaction reduces to an interaction proportional to the square of the hopping term \cite{gotz_phases_2024}
\begin{equation}
  - \frac{g^2}{2k} \sum_{i,\delta} \left( \sum_{\sigma} \fd_{i,\sigma} e^{i\hat{A}_{i,\delta}} \f_{i+\delta,\sigma} + \hc \right)^2.
\end{equation}

\section{Results and discussion}
\label{sec:results}

Before starting the discussion of the QMC results, we first introduce the physical observables we calculate to characterise symmetry-broken phases, phase transitions and calculate spectral excitations. To detect the various phases we measure the imaginary-time structure factor 
\begin{equation}
  S_{O}(\vec{q}, \tau) = \frac{1}{N} \sum_{i,j} e^{i\vec{q}\cdot(\vec{r}_i - \vec{r}_j)} (\langle \hat{O}_i(\tau) \hat{O}_j \rangle - \langle \hat{O}_i \rangle \langle \hat{O}_j \rangle),
\end{equation}
for some local observable \(\hat{O}_i\). We also measure the susceptibility defined as 
\begin{equation}
  \chi_{O}(\vec{q}) = \int_0^\beta d\tau S_{O}(\vec{q}, \tau).
\end{equation}
Moreover, to characterise phase boundaries, we compute the correlation ratio 
\begin{equation} \label{eq:correlation_ratio}
  r_{O} = 1 - \frac{S_{O}(\vec{Q} + \delta\vec{q})}{S_{O}(\vec{Q})},
\end{equation}
where \(\left|\delta\vec{q}\right| = 2\pi/L\) and \(\vec{Q}\) is the ordering wave vector for observable \(\hat{O}\). In the thermodynamic limit, \(r_{O}\) converges to zero (unity) in an disordered (ordered) phase. Furthermore, it is an RG invariant quantity, such that in a critical point we expect \(r_{O}(L, J) = f(L^z/\beta, (J - J_c)L^{1/\nu}, L^{-\omega})\), where \(z\) is the dynamical exponent, \(J_c\) is the value of the critical coupling, \(\nu\) is the correlation length and \(\omega\) is the leading correction to the scaling exponent. As we have Lorentz symmetry, \(z = 1\), we set \(\beta = L\), such that \(r_{O}\) curves of different system sizes cross at the critical point. 

We will be concerned with detecting AFM and VBS phases. The AFM order parameter is defined as 
\begin{equation}
  \vec{\hat{O}}_i^{\text{AFM}} = \vec{\hat{S}}_{i} = \frac{1}{2} \fd_{i,\sigma} \vec{\sigma}_{\sigma, \sigma'} \f_{i,\sigma'},
\end{equation}
where \(\alpha,\beta = 1, 2\), with ordering wave vector at \(\vec{\mathrm{M}} = (\pi,\pi)\). The VBS order parameter is defined as 
\begin{equation}
  \hat{O}_{i,\delta}^{\text{VBS}} = \vec{\hat{S}}_{i} \cdot \vec{\hat{S}}_{i+\delta},
\end{equation}
where \(\delta = x, y\) and the ordering wave vector is \(\vec{\mathrm{X}} = (\pi, 0)\) or \(\vec{\mathrm{Y}} = (0, \pi)\), respectively. To detect the Gauss sectors gap, we measure the susceptibility of the Gauss operators:
\begin{equation}
  \hat{Q}_i = - \sum_{\delta} (\hat{E}_{i,\delta} - \hat{E}_{i,-\delta}) + \sum_\sigma \fd_{i,\sigma} \f_{i,\sigma} - 1.
\end{equation}

To study the dynamics, we calculate imaginary-time displaced observables and use the stochastic analytical continuation (SAC) method \cite{sandvik_stochastic_1998,beach_identifying_2004,shao_progress_2023} to perform the analytical continuation to real frequencies. The Green's function is defined as 
\begin{equation}
  G(\vec{k}, \tau) = \sum_\sigma \langle \fd_{\vec{k},\sigma}(\tau) \f_{\vec{k},\sigma}(0) \rangle.
\end{equation}
We can obtain the single particle spectral function \(A(\vec{k}, \omega)\) through the SAC method:
\begin{equation}
  G(\vec{k}, \tau) = \frac{1}{\pi} \int d\omega \frac{e^{-\tau \omega}}{1 + e^{-\beta \omega}} A(\vec{k}, \omega).
\end{equation} 
We also measure the local or momentum integrated Green's function 
\begin{equation}
  G_0(\tau) = \frac{1}{N} \sum_{i,\sigma} \langle \fd_{i,\sigma}(\tau) \f_{i,\sigma}(0) \rangle.
\end{equation}
Finally, we calculate the spin and dimer dynamical structure factors \( S_O(\vec{q}, \omega)\) through the SAC method: 
\begin{equation}
  S_O(\vec{q}, \tau) = \frac{1}{\pi} \int d\omega e^{-\tau \omega} S_O(\vec{q}, \omega).
\end{equation}
Here \(S_{O}(\vec{q}, \omega) = \chi''_{O}(\vec{q}, \omega) / (1 - e^{-\beta \omega})\), with \(\chi''_{O}(\vec{q}, \omega)\) being the imaginary part of the dynamical susceptibility.

\subsection{Phase diagram}

We now summarise the QMC results for the Hamiltonian in Eq.~\eqref{eq:hamiltonian} through the ground-state phase diagram shown in Fig.~\ref{fig:phase_diagram}. Here we set \(K = t = 1\) as the energy unit and set the phonon frequency \(\omega_0 = 2\). The phase diagram is obtained by varying the electric field coupling \(J\), which controls the gauge field fluctuations, and spinon-phonon coupling \(g\). Together, these parameters allow us to tune between the OSM, AFM, and VBS phases. 

At \(g = 0\), the OSM at small, but finite, \(J\) is stable against weak gauge field fluctuations, and we refer to the surrounding regime as the OSM phase. The characteristic feature of this phase  is a non-activated behaviour of the susceptibility associated to the fluctuations of Gauss's law \( \sum_{i} \hat{Q}_i \). This is a stable phase that extends to finite values of \(J\). The instability of the OSM phase follows two possible scenarios. If the deconfined gauge theory, with dynamically imposed Gauss's law, is stable, an intermediate DSL phase should separate the OSM and AFM phases \cite{xu_monte_2019}. Alternatively, if monopoles proliferate, the system should undergo a single transition from the OSM phase to AFM long-range order \cite{hermele_stability_2004,song_unifying_2019}. Our results support the latter scenario: increasing \(J\) drives a direct OSM-AFM transition at which Gauss's law is dynamically imposed. Then \(J\) controls the condensation of the monopoles carrying AFM quantum numbers in the low-energy theory \cite{alicea_monopole_2008,hermele_algebraic_2005,song_unifying_2019,song_spinon_2020}. 

At small \(J\), increasing \(g\) instead drives a transition from the OSM phase to a VBS phase with ordering wave vector \((\pi,0)\), or equivalently \((0,\pi)\) by lattice rotation symmetry. Gauss's law is also dynamically imposed at this transition, suggesting that the spinon-phonon coupling controls the condensation of monopoles carrying VBS quantum numbers. The OSM-AFM and OSM-VBS boundaries meet the direct AFM-VBS boundary at a multicritical point, producing the three phase structure shown in Fig.~\ref{fig:phase_diagram}.

For sufficiently large \(J\), fine tuning \(g\) to the boundary value \(g_c(J)\) takes the system directly from AFM to VBS order. This point is fine tuned in the usual critical sense: detuning \(g - g_c\) in one direction selects the AFM phase, whereas detuning it in the other direction selects the VBS phase. Despite the unrelated broken symmetries of these two phases, the finite-size evolution of our correlation ratio data is consistent with a single continuous critical point, with no resolved intervening phase or coexistence regime. Our data therefore support a second order AFM-VBS transition, which we identify as a DQCP \cite{senthil_quantum_2004,senthil_deconfined_2004,wang_deconfined_2017}. In this interpretation, the transition lies beyond the conventional LGW paradigm and is governed by fractionalised degrees of freedom coupled to an emergent gauge field for which an enlarged SO(5) symmetry is spontaneously broken to SO(3) \(\times\) O(2).

Below we show our QMC results which support our claims and discuss some exotic aspects of the phases and phase transitions.

\subsection{OSM-AFM transition and OSM phase}

We first focus on the case \(g=0\), for which the Hamiltonian in Eq.~\eqref{eq:hamiltonian} has been extensively studied in Refs.~\cite{xu_monte_2019,janssen_confinement_2020,wang_dynamics_2019,feng_scalable_2026}. Previous studies have discussed the small \(J\) regime in terms of a stable deconfined DSL. Here we revisit this interpretation in light of the stability analysis of the OSM phase presented above. To distinguish the two possibilities, we consider both spin and dimer observables and quantities that probe the energetic separation between different Gauss sectors.

\begin{figure}[t]
  \centering
  \includegraphics[width=.4\textwidth]{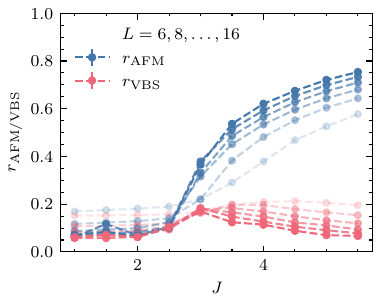}
  \caption{AFM (blue) and VBS (red) correlation ratios as functions of \(J\) at \(g=0\) for \(L=6,8,\ldots,16\). The AFM curves cross near \(J_c \simeq 2.6\). The VBS ratio develops a maximum in the same region. Here we use \(\beta = L\).}
  \label{fig:afm_correlation_ratio}
\end{figure}

Fig.~\ref{fig:afm_correlation_ratio} shows the AFM and VBS correlation ratios as functions of \(J\) and the system size \(L\). The AFM curves exhibit a crossing near \(J_c \simeq 2.6\): below the crossing, \(r_{\text{AFM}}\) decreases with increasing \(L\), whereas above it the ratio grows towards its ordered value. By contrast, \(r_{\text{VBS}}\) remains small and decreases with \(L\) away from the finite-size maximum near the transition. The data therefore identify a direct transition from a phase without AFM or VBS long-range order to an AFM phase, with enhanced VBS correlations at the critical point.

\begin{figure}[t]
  \centering
  \includegraphics[width=.475\textwidth]{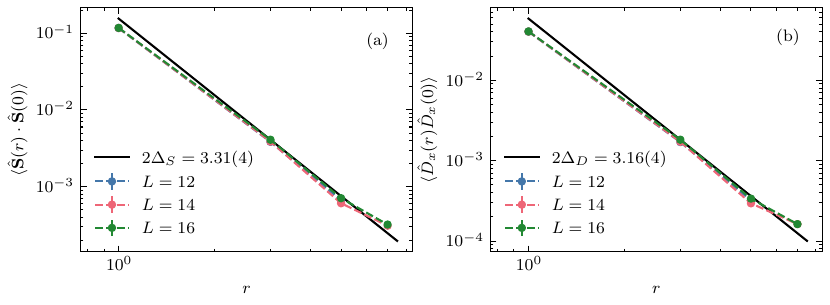}
  \caption{Equal-time real-space (a) spin and (b) dimer correlations at \(J=1\) and \(g=0\). The black lines show algebraic fits, yielding \(2\Delta_S=3.31(4)\) and \(2\Delta_D=3.16(4)\). Here we use \(\beta = L\).}
  \label{fig:real_space_afm_vbs}
\end{figure}

\begin{figure}[t]
  \centering
  \includegraphics[width=.475\textwidth]{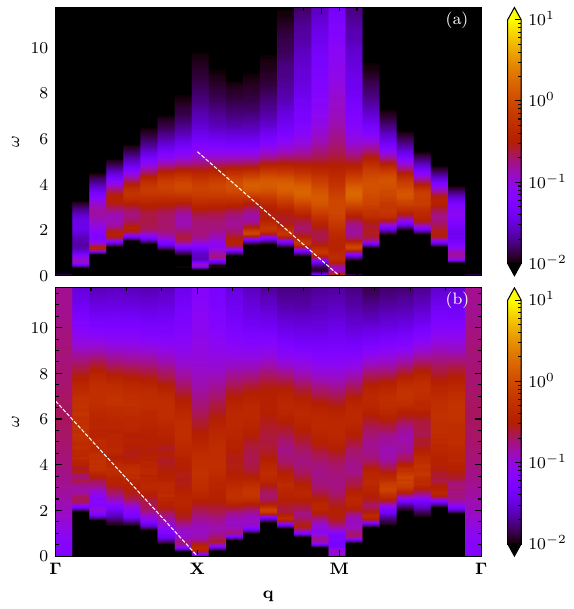}
  \caption{Spin (a) and dimer (b) dynamical structure factors for \(L=16\), \(J=1\), and \(g=0\), obtained by SAC. The dashed lines are guides to the approximately linear low-energy thresholds. Here we use \(\beta = L\).}
  \label{fig:spectral_afm_vbs}
\end{figure}

The correlation ratios characterise the unordered side of the transition. At \(J = 1\), the real-space spin and dimer correlations in Fig.~\ref{fig:real_space_afm_vbs} decay algebraically and show no tendency to saturate with distance. Power-law fits to the finite-size data give \(2\Delta_S = 3.31(4)\) and \(2\Delta_D = 3.16(4)\) for the spin and dimer scaling dimensions, respectively, at \(L = 16\). For non-interacting Dirac fermions in two spatial dimensions, the spin and dimer bilinears have scaling dimension \(\Delta=2\), and hence their correlations decay with \(2\Delta=4\). The fact that the scaling dimensions  of the spin and dimer correlations  are identical reveal important properties of the OSM phase.  It is tempting to write the spin operator in terms of \(c\)-fermions as in Eq.~\eqref{eq:spin_operator_c} and then use the mean-field Hamiltonian of Eq.~\ref{eq:hamiltonian_J0} to compute the dimer correlations. This would however yield a scaling dimension \(\Delta = 4\) in contradiction with the numerical results. The numerical results suggest that the  dimer operator is instead given by \cite{wen_quantum_2002}: 
\begin{equation}
  \ve{\hat{S}}_i \cdot \ve{\hat{S}}_{i+\delta}  \simeq  \hat{f}^{\dagger}_{i,\sigma} \hat{U}_{i,\delta} \hat{f}_{i+\delta,\sigma} + \text{h.c.}
\end{equation} 
The above equation can be justified in terms of symmetries  as well in the realm  of  the U(1) gauge theory of the  Heisenberg model, provided that amplitude fluctuations of the bond variables can be neglected. This form of the dimer  operator  reveals  that,  at least at the mean-field level, it has the same scaling dimension as the spin operator. This is supported  by the numerical results.

Our estimates are substantially below the free Dirac value. This deviation should not, however, be interpreted as conclusive evidence for the existence of a DSL phase as these exponents suffer from strong finite-size effects. Indeed, hybrid Monte Carlo simulations of lattices as large as \(L = 66\) in Ref.~\cite{feng_scalable_2026} find \(2\Delta \simeq 3.8(3)\) for the corresponding fermion-bilinear correlations, much closer to the free Dirac scaling dimensions. The comparison shows that our estimates obtained from \(L = 16\) data can still undergo a substantial drift towards their asymptotic values, indicating slow renormalisation group flow to the OSM fixed point. 

The scaling dimensions are consistent with the Dirac description of the small \(J\) phase and its enlarged low-energy flavour symmetry. The dynamical structure factors in Fig.~\ref{fig:spectral_afm_vbs} provide complementary evidence: the spin and dimer spectra display gapless and linear low-energy dispersion at their respective ordering wave vectors, with comparable velocities \(v_F = 0.67(8)\) and \(v_F = 0.8(2)\), for the spin and dimer order parameters, respectively. Moreover, the low-energy structure of both spectra is similar with gapless modes at the \(\vec{\mathrm{\Gamma}}\), \(\vec{\mathrm{M}}\) and \(\vec{\mathrm{X}}\) points. Thus, the phase at small \(J\) phase retains gapless Dirac like excitations in the spin and dimer channels. These observables alone, however, do not distinguish the OSM phase from a DSL; that distinction is encoded in the structure of the Gauss sectors.

\begin{figure}[t]
  \centering
  \includegraphics[width=.4\textwidth]{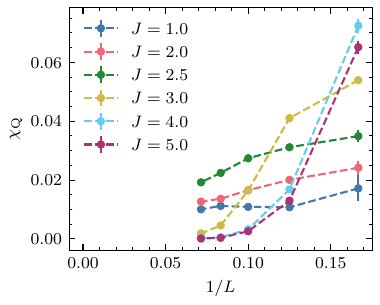}
  \caption{Gauss sector susceptibility \(\chi_Q\) as a function of \(1/L\) and \(J\) at \(g=0\). Here we use \(\beta = L\).}
  \label{fig:gauss_suscep}
\end{figure}

\begin{figure}[t]
  \centering
  \includegraphics[width=.475\textwidth]{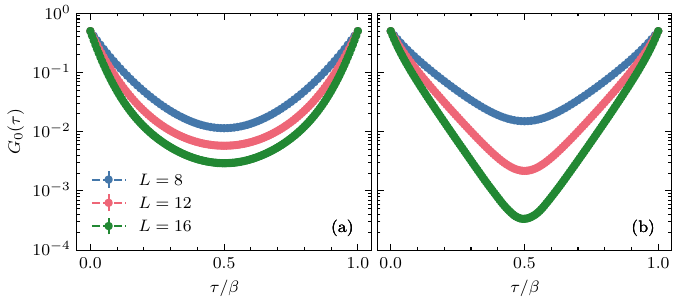}
  \caption{Local imaginary-time Green's function \(G_0(\tau)\) at \(g=0\) in the OSM phase, \(J=1\) (a), and in the AFM phase, \(J=5\) (b). Here we use \(\beta = L\).}
  \label{fig:greens_function}
\end{figure}

The Gauss sector susceptibility, as shown in Fig.~\ref{fig:gauss_suscep}, changes qualitatively across the transition. For \(J \lesssim J_c\), \(\chi_Q\) remains finite as the thermodynamic limit is approached. In particular, \(\chi_Q \) is not activated and for $J=1$ it appears roughly constant with system size. Appendix \ref{app:gauss_suscep} provides  an approximate calculation supporting this observation.

For \(J \gtrsim J_c\), \(\chi_Q \)  is exponentially suppressed with increasing \(L\), \(\chi_Q \sim e^{-\Delta_G / L}\), indicating that fluctuations into sectors with non-zero Gauss charge have a finite gap. The local Green's function \(G_0(\tau)\) in Fig.~\ref{fig:greens_function} supports this interpretation. The insertion of an \(f\)-fermion changes the local Gauss charge as \([\hat{Q}_j, \fd_{i,\sigma}] = \delta_{i,j} \fd_{j,\sigma}\). Between \(J = 1\) and \(J = 5\) we observe two distinct behaviours. At \(J = 1\), \(G_0(\tau)\) decreases slowly with system size, as expected when the sector-changing excitation is gapless. 
Using  the slave boson formulation  introduced in Eq.~\ref{eq:slave_boson}  
\begin{equation}
  G_0(\tau) = \langle \hat{f}^{\dagger}_{0,\sigma}(\tau) \hat{f}_{0,\sigma}(0) \rangle = \langle \hat{c}_{0,\sigma}^{\dagger}(\tau) \hat{b}_0(\tau) \hat{b}_0^\dagger(0) \hat{c}_{0,\sigma}^\dagger(0) \rangle.
\end{equation}
At the mean-field level,   where $\hat{b}$ is a phase,   we  expect   $G_0(\tau)$, after analytical continuation to real time, to show a Dirac semimetallic 
density of states. This is supported by Fig.~\ref{fig:greens_function_spectral}. In contrast, at \(J = 5\), the much stronger suppression around \(\tau = \beta/2\) is consistent with its exponential decay as a function of \(\tau\) and hence with a finite gap between the ground-state Gauss sector and the lowest sector reached by the fermion operator. Taken together, these results indicate that Gauss's law becomes energetically imposed upon entering the AFM phase.  

In summary, our results support the existence of a stable OSM phase at small \(J\) and show no evidence for a DSL extending over a finite interval of \(J\). Nevertheless, the VBS correlations are enhanced near \(J_c\), see Fig.~\ref{fig:afm_correlation_ratio}. The maximum in \(r_{\text{VBS}}\) occurs in the same region as the AFM correlation ratio crossing. This simultaneous enhancement of AFM and VBS correlations may indicate that the critical point has a higher symmetry, i.e. SO(5). On the square lattice, the DSL is known to host such symmetry \cite{song_unifying_2019,song_spinon_2020} which relates the monopole operators associated with the competing orders. Such a DSL exists only at criticality and is unstable to monopoles. Their proliferation confines the spinons and selects the AFM ordered phase. The data is therefore consistent with a direct OSM-AFM transition mediated by an unstable DSL critical point, rather than with an extended DSL phase.

\subsection{Imposing Gauss's law}

So far, we have simulated an unconstrained lattice gauge theory, for which the local Gauss operators \(\hat Q_i\) are conserved but their eigenvalues are not fixed. In the OSM phase, the energy differences between distinct Gauss sectors collapse, so configurations with different sets of local charges \(\{Q_i\}\) remain accessible at low energies. This raises a natural question: what happens if the physical constraint \(Q_i = 0\) is imposed explicitly rather than generated dynamically? 

To impose Gauss's law explicitly, we add an energy penalty to the Hamiltonian:
\begin{equation}
  \hat{H}_\lambda = \hat{H} + \lambda \sum_i \hat{Q}_i^2.
\end{equation} 
Since \([\hat{Q}_i,\hat{H}]=0\), this term shifts the relative energies of the Gauss sectors \(\{Q_i\}\). Taking \(\lambda\to\infty\) projects exactly onto the physical Hilbert space. As derived in Appendix \ref{app:imposing_gauss_law}, this projection introduces a compact temporal gauge field \(a^0_{i,\tau}\) into the path integral formulation. This field acts as a Lagrange multiplier coupled to the local Gauss operator: integrating \(e^{i a^0_{i,\tau}\hat{Q}_i}\) eliminates all contributions with \(Q_i \neq 0\), enforcing the relation between the matter charge and the electric-field divergence at every site and imaginary-time slice. This formulation allows us to examine whether the OSM phase survives when the statistical ensemble is restricted to the physical sector.

\begin{figure}[t]
  \centering
  \includegraphics[width=.4\textwidth]{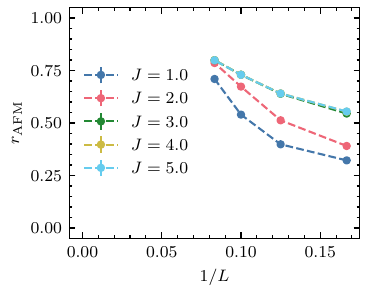}
  \caption{AFM correlation ratio as a function of \(1 / L\) at \(g = 0\) after exactly imposing the local constraint \(Q_i = 0\), with \(\lambda \to \infty\). Here we use \(\beta = L\).}
  \label{fig:gauss_constraint}
\end{figure}

Fig.~\ref{fig:gauss_constraint} shows the AFM correlation ratio of the constrained Hamiltonian \(\hat{H}_{\lambda \to \infty}\). For all values of \(J\) presented, \(r_{\text{AFM}}\) increases as a function of \(L\). In particular, the data at \(J \lesssim J_c\) is compatible with AFM long-range order, even though, when all Gauss sectors are included, in the same regime we have an OSM phase. Therefore, constrained data shows no evidence for either an OSM phase or an intermediate DSL: once Gauss's law is imposed, the physical sector is already AFM ordered.

The comparison between the constrained and unconstrained calculations clarifies the origin of the OSM phase. It is not a disordered state intrinsic to the \(Q_i = 0\) sector. Instead, it relies on the collapse of the energy separation between different Gauss sectors. Adding a penalising term and taking \(\lambda \to \infty\) introduces a temporal gauge field \(a^0\) which, upon integration, exactly projects the theory to the physical sector and resulting in AFM order. More importantly, without this temporal component, and in a regime where Gauss's law is not dynamically imposed, monopoles cannot be defined. As so, the OSM phase is stable in the unconstrained theory. Adding the constraint or increasing \(J\) in the unconstrained model, produces the same selection dynamically: the Gauss sectors become gapless, the \(Q_i = 0\) sector becomes energetically isolated, and AFM order develops due to the relevance of monopole excitation s \cite{song_unifying_2019}.

\subsection{OSM-VBS transition}

We now consider the OSM-VBS transition, in our unconstrained Hamiltonian, by fixing \(J = 1\) and increasing the spinon-phonon coupling \(g\). The bond phonons transform under lattice symmetries as the VBS order parameter and hence couple directly to the monopoles carrying VBS quantum numbers \cite{seifert_spin-peierls_2024,hofmeier_spin-peierls_2024}. Schematically, this coupling can be written as 
\begin{equation}
  S_{\Phi\Phi} \sim - g^2 \int d\tau d\tau' \Phi_a^\dagger(\tau) e^{-\omega_0 |\tau - \tau'|} \Phi_a(\tau'),
\end{equation}
where \(\Phi_{a=x,y}\) denotes the VBS monopoles. The phonon frequency tunes the range, in imaginary time of the effective interaction, whilst increasing \(g\) enhances VBS monopole fluctuations and ultimately drives their condensation, producing VBS long-range order and confining the gauge theory.

\begin{figure}[t]
  \centering
    \includegraphics[width=.475\textwidth]{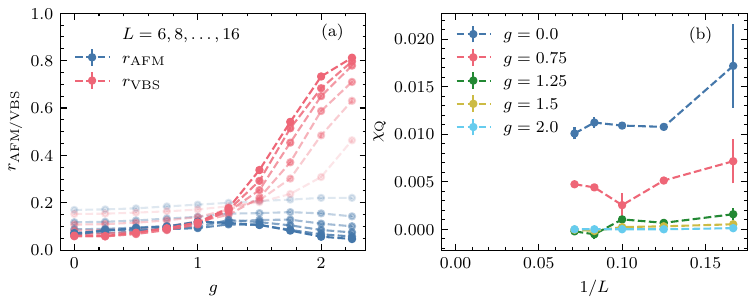}
  \caption{OSM-VBS transition at \(J=1\). (a) AFM and VBS correlation ratios as functions of \(g\) for \(L=6,8,\ldots,16\). The VBS curves cross at \(g_c \simeq 1.1 \text{-} 1.3\), while the AFM curves remain small. (b) Gauss sector \(\chi_Q\) susceptibility as a function of \(1/L\). Here we use \(\beta = L\).}
  \label{fig:higgs_vbs_transition}
\end{figure}

Fig.~\ref{fig:higgs_vbs_transition}(a) shows the AFM and VBS correlation ratios across the transition. For \(g < g_c\), \(r_{\text{VBS}}\) decreases with increasing system size, consistently with the absence of VBS long-range order in the OSM phase. Above \(g_c \simeq 1.2\), \(r_{\text{VBS}}\) grows towards the ordered limit. By contrast, \(r_{\text{AFM}}\) remains small and decreases with system size throughout the transition. In particular, we also observe a small enhancement of the spin correlations at criticality, Fig.~\ref{fig:afm_correlation_ratio}. Furthermore, the Gauss sector susceptibility in Fig.~\ref{fig:higgs_vbs_transition}(b) provides evidence that the transition also changes the structure of the low-energy Hilbert space. As before, in the OSM phase \(g \lesssim 1.2\), \(\chi_Q\) decreases linearly with \(1/L\), whilst in the VBS phase it is strongly suppressed. So, at the transition point Gauss's law becomes energetically imposed, as in the case for the OSM-AFM transition.

Close to the transition point, there is a finite-size enhancement of the spin correlations, suggesting a larger symmetry at criticality. Thus, our data suggests that this transition is also governed by a deconfined DSL which is unstable to confinement due to the proliferation of monopoles. Then, our results support a direct picture in which \(g\) tunes the VBS monopole condensation upon the energetic imposition of Gauss's law.

\subsection{AFM-VBS transition}

We now consider the direct transition between the AFM and VBS phases at large \(J\). We fix \(J = 5\) and use the spinon-phonon coupling \(g\) to tune across the phase boundary. In the compact QED\(_{2+1}\) description, the AFM and VBS order parameters are associated with different components of the monopole multiplet \cite{song_unifying_2019,song_spinon_2020}. The electric field coupling \(J\) favours the monopoles carrying AFM quantum numbers, whereas the phonons couple directly to the VBS monopoles. Reaching the AFM-VBS critical point therefore requires fine tuning \(g\) to a critical value \(g_c(J)\) at which there is no competition between the different monopoles, i.e. the SO(5) symmetric point. For \(g<g_c\), the AFM monopole channel is selected, while for \(g>g_c\), the VBS monopoles condense. Then for \(g \neq g_c\) we have an SO(5) symmetry breaking to SO(3) \(\times\) O(2) \footnote{On the lattice the O(2) symmetry is broken to C\(_4\)}.

A direct continuous transition between AFM and VBS phases lies beyond the LGW paradigm because the two phases break unrelated symmetries \cite{senthil_quantum_2004}. The standard bosonic description of this deconfined quantum critical point is the non-compact CP\(^{1}\) theory, in which fractionalised bosonic spinons are coupled to an emergent U(1) gauge field \cite{senthil_quantum_2004,senthil_deconfined_2004}. Complementary fermionic descriptions organise the three AFM and two VBS components into a five-component vector and with an emergent SO(5) symmetry \cite{senthil_competing_2006,tanaka_many-body_2005,wang_deconfined_2017,nahum_emergent_2015}. These bosonic and fermionic formulations are conjectured to provide dual descriptions of the same critical point, i.e. the DQCP.

\begin{figure}[t]
  \centering
    \includegraphics[width=.475\textwidth]{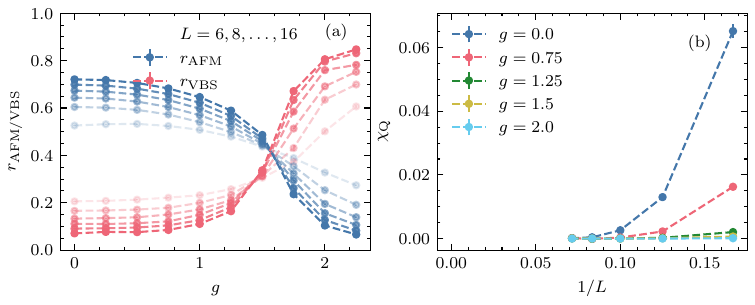}
  \caption{AFM-VBS transition at \(J=5\). (a) AFM and VBS correlation ratios as functions of \(g\) for \(L=6,8,\ldots,16\). The opposite finite-size flows of the two ratios locate a direct transition at \(g_c \simeq 1.5 \text{-} 1.7\). (b) Gauss-sector susceptibility \(\chi_Q\) versus \(1/L\). Here we use \(\beta = L\).}
  \label{fig:correlation_ratio_dqcp}
\end{figure}

Fig.~\ref{fig:correlation_ratio_dqcp}(a) shows the AFM and VBS correlation ratios across the transition. At small \(g\), \(r_{\text{AFM}}\) increases with system size and approaches the ordered limit. On the other hand, at large \(g\), \(r_{\text{VBS}}\) grows towards the VBS-ordered limit. The two ratios become comparable in the interval \(g_c \simeq 1.5\). Within our finite-size data, we do not resolve an intermediate disordered phase or a finite coexistence region, indicating that the AFM and VBS phases meet at a single direct transition. Furthermore, the Gauss sector susceptibility in Fig.~\ref{fig:correlation_ratio_dqcp}(b) tends towards zero with increasing system size for every value of \(g\) shown, including \(g=1.5\) in the critical region. Therefore, Gauss's law is already energetically imposed in the AFM phase and remains imposed through the transition and into the VBS phase. So, the AFM-VBS transition takes place entirely within the physical Hilbert space and corresponds to changing which component of the monopole multiplet condenses.

\begin{figure}[t]
  \centering
  \includegraphics[width=.475\textwidth]{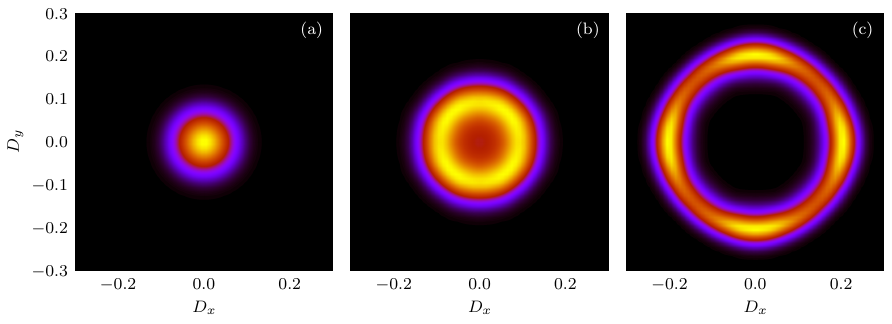}
  \caption{Normalised histograms of the two component VBS order parameter \((D_x,D_y)\) for \(L=14\) and \(J=3\): (a) \(g=0.75\), in the AFM phase; (b) \(g=1.5\), near the AFM-VBS transition; and (c) \(g=2.25\), in the VBS phase. Here we use \(\beta = L\).}
  \label{fig:histogram_vbs}
\end{figure}

The evolution of the VBS order parameter distribution is shown in Fig.~\ref{fig:histogram_vbs}. In the AFM phase at \(g = 1.25\), the distribution is centred at the origin, consistently with the absence of VBS order. In the VBS phase \(g = 2.25\), we observe a four peak structure corresponding to the two VBS orders related by a C\(_4\) rotation. At criticality, \(g = 1.75\), the distribution broadens into a ring. The angular dependence associated with the microscopic \(C_4\) symmetry is not resolved, indicating that the fourfold anisotropy is strongly suppressed and that the VBS order parameter acquires an approximate U(1) symmetry. This emergent U(1) symmetry is a characteristic signature of deconfined quantum criticality and can be understood as a consequence of the irrelevance of the lattice monopole anisotropy at the critical point \cite{senthil_quantum_2004,senthil_deconfined_2004,sandvik_evidence_2007,gotz_phases_2024,gotz_tuning_2024}. 

\begin{figure*}[t]
  \centering
  \includegraphics[width=1\textwidth]{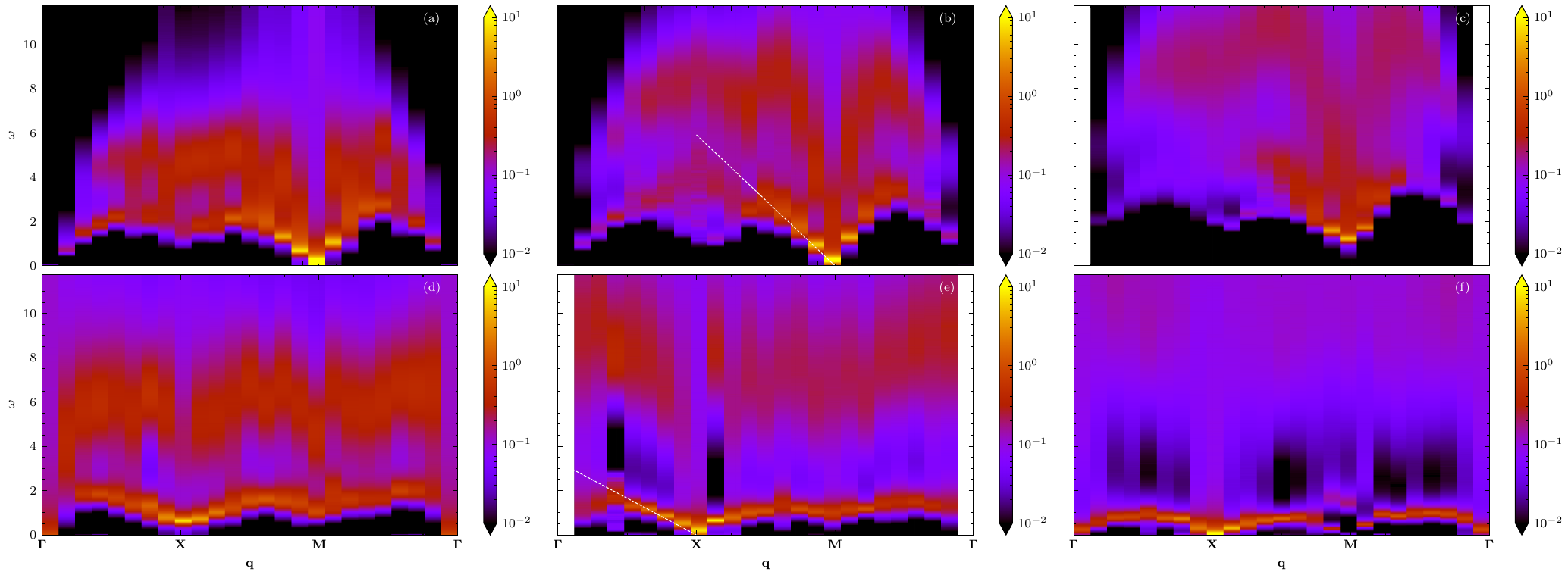}
  \caption{Spin (top row) and dimer (bottom row) dynamical structure factors for \(L=16\) and \(J=5\). The columns correspond to \(g=1.0\) in the AFM phase, \(g=1.5\) near the AFM--VBS transition, and \(g=2.0\) in the VBS phase. The dashed lines are guides to the low-energy dispersions. Here we use \(\beta = L\).}
  \label{fig:spectral_dqcp}
\end{figure*}

The dynamical structure factors in Fig.~\ref{fig:spectral_dqcp} show how the low-energy spectral weight is transferred between the two competing orders. In the AFM phase at \(g = 1\), the dominant low-energy spin excitation is located at the AFM ordering wave vector \(\vec{\mathrm{M}}\), while the dimer response at \(\vec{\mathrm{X}}\) remains comparatively less singular. Conversely, in the VBS phase \(g = 2\), the spin response is gapped whilst the dimer has a dominant peak at \(\vec{\mathrm{X}}\). Near the transition at \(g = 1.5\), both the spin excitations at \(\mathbf M\) and the dimer excitations at \(\vec{\mathrm{X}}\) soften. This simultaneous presence of low-energy spin and dimer excitations is consistent with critical AFM and VBS fluctuations. Together with the emergent U(1) symmetry of the VBS order parameter, suggests that the two order parameters become part of a five-component vector which at criticality has an enhanced SO(5) symmetry. 

Taken together, the direct onset of AFM and VBS order, the absence of an intervening phase, the suppression of the VBS anisotropy, and the simultaneous softening of the spin and dimer spectra support the picture of a continuous AFM-VBS transition, i.e. a deconfined quantum critical point, within the accessible system sizes. In the monopole language, \(g\) fine tunes the system between two confined phases distinguished by which monopole component condenses. At \(g = g_c\), neither ordering tendency dominates and both become critical. Even though the data does not show any clear evidence of a strong first-order transition, a weak first-order transition cannot be excluded without simulations at larger sizes.

\section{Conclusion}
\label{sec:conclusion}

In this work, we used exact AF-QMC simulations to investigate an unconstrained compact U(1) lattice gauge theory coupled to bond phonons with \(N_f = 2\) fermion flavours. With our simulations we map the ground state phase diagram as a function of electric field strength \(J\) and spinon-phonon coupling \(g\). The results are summarised in Fig.~\ref{fig:phase_diagram}. In the unconstrained theory, the local Gauss operators are conserved but their eigenvalues are not fixed. Gauss sectors are degenerate at  \(J  \rightarrow 0\); our perturbative and numerical results support an OSM phase at finite \(J\), in which the Gauss sector gap vanishes while spin and dimer operators show signatures of Dirac physics. Exactly penalising local Gauss sector excitations, produces AFM order in the same parameter region. Thus we find no evidence for a stable DSL phase at \(N_f = 2\). The temporal component of the gauge field acts as a Lagrange multiplier enforcing Gauss's law. Since Gauss's law is not imposed dynamically in the OSM, the temporal component of the gauge field is absent. As a consequence, the very definition of a monopole is invalidated. We understand this to be the reason why the OSM is a stable phase of matter in compact lattice QED\(_{2+1}\).

The transitions out of the OSM phase are associated with the dynamical generation of Gauss's law and concomitant monopole excitations. Increasing \(J\) or \(g\) drives direct OSM-AFM or OSM-VBS confinement transitions, which we interpret in terms of the condensation of the corresponding monopoles. In both cases, symmetry breaking coincides with the energetic imposition of Gauss's law. Enhanced VBS (AFM) fluctuations on the OSM-AFM (OSM-VBS) transition suggests a higher symmetry and deconfinement at criticality. At larger \(J\), where Gauss's law is already energetically imposed, by tuning \(g\) we find a direct AFM-VBS transition with no coexistence regime. The emergent U(1) symmetry of the VBS order parameter and the simultaneous softening of spin and dimer excitations at the critical point, provides evidence for a DQCP. Nevertheless, a weak first-order transition cannot be excluded without simulations on larger lattices.

The point \(J = 0\) is singular, i.e., the gauge field becomes static, and selecting a fixed background breaks local U(1) invariance from the effective fermionic Hamiltonian. At finite \(J\), the signatures of enlarged symmetry at the OSM–AFM and OSM–VBS transitions suggest that a DSL may be realised at criticality. Relevant monopole excitations then drive confinement and select the adjacent AFM or VBS order. The direct AFM–VBS transition is likewise consistent with a DQCP, at which fractionalised excitations become deconfined. Together, these results suggest that deconfinement is a common feature of all three transitions in our unconstrained lattice QED model, even though we find no evidence for a stable DSL phase extending over a finite parameter region.

\begin{acknowledgments}
  We thank discussions with C. Chen, N. Costa, I. Herbut, L. Janssen, Z. Y. Meng, G. Pan and J. Willsher. 
  We  gratefully acknowledge the Gauss Centre for Supercomputing e.V. for funding this project by providing computing time on the GCS Supercomputer SUPERMUC-NG at Leibniz Supercomputing,   (project number pn73xu) as  well  as  the scientific support and HPC resources provided by the Erlangen National High Performance Computing Center (NHR@FAU) of the Friedrich-Alexander-Universit\"at Erlangen-N\"urnberg (FAU) under the NHR project b133ae. NHR funding is provided by federal and Bavarian state authorities. NHR@FAU hardware is partially funded by the German Research Foundation (DFG) -- 440719683. 
  J.C.I. thanks the DFG for financial support under the AS 120/19-1 grant (Project number, 530989922).
  F.F.A.  acknowledges financial support by the Deutsche Forschungsgemeinschaft (DFG, German Research Foundation) through the W\"urzburg-Dresden Cluster of Excellence {\it ctd.qmat} -- Complexity, Topology and Dynamics in Quantum Matter (EXC 2147, Project No.~390858490). 
  Part of this work was finalized during a  workshop on Correlated Flat Bands, Peyresq, France, 26 July to 2 August 2026, funded by  the  Julian Schwinger Foundation.  We thank this foundation for financial support.
\end{acknowledgments}

\section*{Data availability}
The data that support the findings of this article is available upon request.

\appendix

\begin{widetext}
\section{Auxiliary-field quantum Monte Carlo formulation and Metropolis assisted Langevin algorithm}
\label{app:af_qmc}

In this section we present the auxiliary-field quantum Monte Carlo formulation of the compact U(1) lattice gauge theory coupled to phonons described in Hamiltonian \eqref{eq:hamiltonian}. For this we start by writing the gauge and phonon Hamiltonians in the path integral formalism. Then we integrate the fermions out and get a purely bosonic action for which we can sample with the Monte Carlo method. 

The gauge Hamiltonian is given by:
\begin{equation}
  \hat{H}_{\text{gauge}} = \frac{J}{4} \sum_{i,\delta} \hat{E}^2_{i,\delta} + K \sum_\square \cos(\text{curl} \hat{A}_\square).
\end{equation}
Here \([\hat{A}_{i, \delta}, \hat{E}_{i^\prime, \delta^\prime}] = i \delta_{i,i^\prime} \delta_{\delta, \delta^\prime}\) and \(\text{curl} \hat{A}_\square = \hat{A}_{i,x} + \hat{A}_{i+x,y} - \hat{A}_{i+y,x} - \hat{A}_{i, y}\) is the flux over a plaquette on the square lattice. For each bond, we define the eigenstates of \(\hat{A}\) and \(\hat{E}\) as 
\begin{equation}
  \hat{A} \ket{a} = a \ket{a}, \quad \hat{E} \ket{n} = n \ket{n}.
\end{equation}
As we are interested in the case where the gauge fields are compact, i.e. the Hilbert space of \(\hat{A}\) is \(L^2([0, 2\pi])\). Thus \(a \in [0, 2\pi]\) and \(n \in \mathbb{Z}\) is an integer. Here \(\braket{a}{a^\prime} = \delta(a - a^\prime)\) and \(\braket{n}{n^\prime} = \delta_{n, n^\prime}\). In the eigenbasis of \(\hat{A}\), the electric field is given by \(\hat{E} \to - i \frac{\partial}{\partial a}\). The overlap between the two eigenstates is then given by \(\braket{a}{n} = e^{i a n}\) and the completeness relations are \(\int_0^{2\pi} da \ket{a}\bra{a} = \frac{1}{2\pi} \sum_n \ket{n}\bra{n} = 1\). Now we calculate the path integral of the gauge Hamiltonian \(\hat{H}_{\text{gauge}}\). We start with a Trotter discretisation of the imaginary time propagator into  $L_\tau$ time slices of size $\Delta\tau$ ($L_\tau \Delta\tau = \beta$) and perform a path integral using the eigenstates of the $\hat{A}$ operator:
\begin{equation}
  Z_{\text{gauge}} = \text{Tr} \left[ e^{-\beta\hat{H}_{\text{gauge}}} \right] = \int \D{a} \prod_{\tau = 1}^{L_\tau} \bra{a_{\tau+1}} e^{-\Delta\tau \hat{H}_{\text{gauge}}} \ket{a_\tau},
\end{equation}
with the condition that $\ket{a_1} = \ket{a_{L_\tau + 1}}$ and \(\beta = 1/T\). Now we focus on computing the matrix element 
\begin{equation}
  \bra{a_{\tau+1}} e^{-\Delta\tau \hat{H}_{\text{gauge}}} \ket{a_\tau} =  e^{-\Delta\tau K \sum_\square \cos(\text{curl} a_{\square, \tau})} \prod_{i,\delta} \bra{a_{i,\delta, \tau+1}} e^{-\Delta\tau \frac{J}{4} \hat{E}^2_{i,\delta}} \ket{a_{i,\delta,\tau}}.
\end{equation}
Inserting an identity and focusing on the element of the product, we have 
\begin{equation}
  \bra{a_{\tau+1}} e^{-\Delta\tau \frac{J}{4} \hat{E}^2} \ket{a_{\tau}} = \frac{1}{2\pi} \sum_{n} \braket{a_{\tau+1}}{n} \bra{n} e^{-\Delta\tau \frac{J}{4} \hat{E}^2} \ket{a_{\tau}} = \frac{1}{2\pi} \sum_{n} e^{-\Delta\tau \frac{J}{4} n^2 + i n (a_{\tau+1} - a_{\tau})}.
\end{equation}
The summation can be solved using the Poisson summation formula:
\begin{equation}
  \frac{1}{2\pi} \sum_{n} e^{-\Delta\tau \frac{J}{4} n^2 + i n (a_{\tau+1} - a_{\tau})} \sim \sum_{k} e^{- \frac{(a_{\tau+1} - a_{\tau} - 2\pi k)^2}{4J \Delta\tau}},
\end{equation}
where \(k \in \mathbb{Z}\). Using the Villain approximation, the gauge partition function is given by \(Z_{\text{gauge}}  = \int \D{a} e^{-S_{\text{gauge}}}\) and the gauge action
\begin{equation}
  S_{\text{gauge}} = \sum_{\tau = 1}^{L_\tau} \left[ \frac{2}{J\Delta\tau} \sum_{i,\delta} (1 - \cos(a_{i,\delta,\tau+1} - a_{i,\delta,\tau})) + \Delta\tau K \sum_{\square} \cos(\text{curl} a_{\square,\tau}) \right].
\end{equation}

The phonon Hamiltonian is given by:
\begin{equation}
  \hat{H}_{\text{phonon}} = \sum_{i,\delta} \left(\frac{\hat{P}_{i,\delta}^2}{2m} + \frac{k}{2} \hat{X}_{i,\delta}^2\right). 
\end{equation}
Here \([\hat{X}_{i,\delta}, \hat{P}_{i^\prime, \delta^\prime}] = i \delta_{i,i^\prime} \delta_{\delta, \delta^\prime}\). For each bond we define the eigenstates of \(\hat{X}\) and \(\hat{P}\) as 
\begin{equation}
  \hat{X} \ket{x} = x \ket{x}, \quad \hat{P} \ket{p} = p \ket{p},
\end{equation}
where \(x, p \in \mathbb{R}\). Here \(\braket{x}{x^\prime} = \delta(x - x^\prime)\) and similar for \(\hat{P}\). In the eigenbasis of \(\hat{X}\), the momentum operator is given by \(\hat{P} \to - i \frac{\partial}{\partial x}\). The overlap between the two eigenstates is then given by \(\braket{x}{p} = e^{i x p}\) and the completeness relations are \(\int dx \ket{x}\bra{x} = \frac{1}{2\pi} \int dp \ket{p}\bra{p} = 1\). Now we calculate the path integral of the phonon Hamiltonian \(\hat{H}_{\text{phonon}}\). We use the same Trotter discretisation and perform a path integral using the eigenstates of the $\hat{X}$ operator:
\begin{equation}
  Z_{\text{phonon}} = \text{Tr} \left[ e^{-\beta\hat{H}_{\text{phonon}}} \right] = \int \D{x} \prod_{\tau = 1}^{L_\tau} \bra{x_{\tau+1}} e^{-\Delta\tau \hat{H}_{\text{phonon}}} \ket{x_\tau},
\end{equation}
with the condition that $\ket{x_1} = \ket{x_{L_\tau + 1}}$. Now we focus on computing the matrix element 
\begin{equation}
  \bra{x_{\tau+1}} e^{-\Delta\tau \hat{H}_{\text{phonon}}} \ket{x_\tau} =  e^{-\Delta\tau \frac{k}{2} \sum_{i,\delta} x_{i,\delta,\tau}^2} \prod_{i,\delta} \bra{x_{i,\delta, \tau+1}} e^{-\Delta\tau \hat{P}^2_{i,\delta} / (2m)} \ket{x_{i,\delta,\tau}}.
\end{equation}
Inserting an identity and focusing on the element of the product, we have 
\begin{equation}
  \bra{x_{\tau+1}} e^{-\Delta\tau \hat{P}^2_{i,\delta} / (2m)} \ket{x_{\tau}} = \frac{1}{2\pi} \int dp \braket{x_{\tau+1}}{p} \bra{p} e^{-\Delta\tau \hat{P}^2_{i,\delta} / (2m)} \ket{x_{\tau}} = \frac{1}{2\pi} \int dp e^{-\Delta\tau p^2 / (2m) + i p (x_{\tau+1} - x_{\tau})}.
\end{equation}
Doing a Gaussian integration, the phonon partition function is given by \(Z_{\text{phonon}}  = \int \D{a} e^{-S_{\text{phonon}}}\) and the phonon action
\begin{equation} \label{eq:phonon_action}
  S_{\text{phonon}} = \sum_{\tau = 1}^{L_\tau} \sum_{i,\delta} \left[ \frac{m}{2\Delta\tau} (x_{i,\delta,\tau+1} - x_{i,\delta,\tau})^2 + \frac{\Delta\tau k}{2} x_{i,\delta, \tau}^2 \right].
\end{equation}

Now we focus on the full lattice gauge theory Hamiltonian with fermions degrees of freedom, equation \eqref{eq:hamiltonian}. After an imaginary time discretisation, the partition function is given by 
\begin{equation}
  Z = \text{Tr}\left[e^{-\Delta\tau (\hat{H}_t + \hat{H}_{\text{gauge}} + \hat{H}_{\text{phonon}})}\right]^{L_\tau},
\end{equation}
where 
\begin{equation}
  \hat{H}_t = \sum_{i, \delta, \sigma} (-t + g \hat{X}_{i,\delta}) \left(\fd_{i, \sigma} e^{i \hat{A}_{i,\delta}} \f_{i+\delta, \sigma} + \text{h.c.}\right) = \sum_{i,\delta} \hat{K}_{i,\delta}(\hat{X}_{i,\delta}, \hat{A}_{i,\delta}), 
\end{equation}
and \(\hat{H}_{\text{gauge}}\) and \(\hat{H}_{\text{phonon}}\) are the gauge and phonon Hamiltonians, respectively. We use a non-symmetric Trotter decomposition 
\begin{equation}
  e^{-\Delta\tau (\hat{H}_t + \hat{H}_{\text{gauge}} + \hat{H}_{\text{phonon}})} = e^{-\Delta\tau \hat{H}_t} e^{-\Delta\tau \hat{H}_{\text{gauge}}} e^{-\Delta\tau \hat{H}_{\text{phonon}}}  + \mathcal{O}((\Delta\tau)^2).
\end{equation}
As the Hamiltonian is bilinear in fermions, we can integrate them out. We then arrive at
\begin{equation}
  Z = \int \D{a,x}  e^{-S_{\text{phonon}} - S_{\text{gauge}}} \det\left(1 + B(\beta, 0)\right)
\end{equation}
where the fermion determinant is given by 
\begin{equation} 
  B(\tau_2, \tau_1) = \prod_{\tau=\tau_1 + \Delta\tau}^{\tau_2} \prod_{i,\delta} e^{- \Delta\tau K_{i,\delta}(x_{i,\delta,\tau}, a_{i, \delta, \tau})}
\end{equation}
This action can be sampled using the auxiliary-field quantum Monte Carlo method where the configuration space is the set \(C = \{x_{i,\delta,\tau}; a_{i,\delta,\tau} \}\) with two continuous fields per bond and time slice.

To efficiently sample the action, we employ the Metropolis assisted Langevin algorithm (MALA) \cite{roberts_exponential_1996}. MALA is a Markov chain Monte Carlo method that combines stochastic Langevin dynamics with a Metropolis accept/reject move. In contrast to purely Metropolis updates, where configurations are proposed independently of the local structure of the action, MALA includes gradient information from the action into the proposal step. The resulting term guides the sampling towards regions of higher statistical weight, significantly improving autocorrelation times. More specifically, MALA constructs proposals of the phonon and gauge fields through a discretised version of the Langevin equation. The deterministic component is the force associated with the action, while the stochastic Gaussian noise ensures ergodicity. Each proposal is then subjected to a Metropolis accept/reject step.

The action we want to sample is given by 
\begin{equation}
  S = S_{\text{phonon}}(\{x\}) + S_{\text{gauge}}(\{a\}) - \ln \text{Tr} \left(\det(1 + B(\beta, 0))\right).
\end{equation}
The new phonon and gauge fields are proposed according to 
\begin{gather}
  x_{b,\tau}^{t+1} = x_{b,\tau}^t - \delta t_{\text{p}} \frac{\partial S}{\partial x_{b,\tau}^t} + \sqrt{2 \delta t_{\text{p}}} \eta(0, 1), \\
  a_{b,\tau}^{t+1} = a_{b,\tau}^t - \delta t_{\text{g}} \frac{\partial S}{\partial a_{b,\tau}^t} + \sqrt{2 \delta t_{\text{g}}} \eta(0, 1). 
\end{gather}
Here \(t\) is the Langevin time, \(\delta t\) is the Langevin time step and \(\eta(0, 1)\) is a random normally distributed variable with mean 0 and variance 1. We propose a move for the phonon and gauge fields are one bond and time slice, simultaneously. The Metropolis acceptance criteria for an update \(\phi_{\text{old}} = \{x_{b,\tau}^t, a_{b,\tau}^t\}\) to \(\phi_{\text{new}} = \{x_{b,\tau}^{t+1}, a_{b,\tau}^{t+1}\}\) is given by 
\begin{equation}
  P(\phi_{\text{old}} \to \phi_{\text{new}}) = \text{min} \left(1, \frac{W(\phi_{\text{old}}) Q(\phi_{\text{old}}, \phi_{\text{new}})}{W(\phi_{\text{new}})Q(\phi_{\text{new}}, \phi_{\text{old}})}\right),
\end{equation}
where 
\begin{equation}
  Q(\phi_{\text{old}}, \phi_{\text{new}}) = \exp\left[-\frac{1}{4 \delta t_{\text{p}}} \left(x_{b,\tau}^{t+1} - x_{b,\tau}^{t} + \delta t_{\text{p}} \frac{\partial S}{\partial x_{b,\tau}^t}\right)^2 \right]  \exp\left[ -\frac{1}{4 \delta t_{\text{g}}} \left(a_{b,\tau}^{t+1} - a_{b,\tau}^{t} + \delta t_{\text{g}} \frac{\partial S}{\partial a_{b,\tau}^t}\right)^2 \right]
\end{equation}
is the transition probability density from state \(\phi_{\text{old}}\) to \(\phi_{\text{new}}\).

To compute the forces, we define the Green's function at time slice \(\tau\) and the local Green's function at bond \(b\) at time slice \(\tau\) as 
\begin{gather}
  G^{-1}(\tau) = 1 + B(\tau, 0) B(\beta, \tau), \\
  G_b^{-1}(\tau) = 1 + B_{b,0}(\tau, 0) B_{N_b, b}(\beta, \tau),
\end{gather}
where \(N_b\) is the number of bonds and 
\begin{gather}
  B_{b, 0}(\tau, 0) = \prod_{b^\prime = 1}^{b} e^{-\Delta\tau K_b(x_{b^\prime,\tau + 1}, a_{b^\prime,\tau + 1})} B(\tau + \Delta\tau, 0), \\
  B_{N_b, b}(\beta, \tau) = B(\beta, \tau + \Delta\tau) \prod_{b^\prime = b+1}^{N_b} e^{-\Delta\tau K_b(x_{b^\prime,\tau + 1}, a_{b^\prime,\tau + 1})}.
\end{gather}
Then, 
\begin{multline}
  \frac{\partial S}{\partial x_{b,\tau}^t} = \frac{\partial S_{\text{phonon}}}{\partial x_{b,\tau}^t} - \frac{\partial}{\partial x_{b,\tau}^t} \ln \text{Tr} \left(\det(1 + B(\beta, 0))\right) \\
  = \frac{\partial S_{\text{phonon}}}{\partial x_{b,\tau}^t} - \text{Tr} \left( e^{\Delta\tau K_{b}(x_{b,\tau}, a_{b,\tau})} \frac{\partial}{\partial x_{b,\tau}^t}  e^{-\Delta\tau K_{b}(x_{b,\tau}, a_{b,\tau})} (1 - G_b(\tau)) \right).
\end{multline}
Similarly,
\begin{equation}
  \frac{\partial S}{\partial a_{b,\tau}^t} = \frac{\partial S_{\text{gauge}}}{\partial a_{b,\tau}^t} - \text{Tr} \left( e^{\Delta\tau K_{b}(x_{b,\tau}, a_{b,\tau})} \frac{\partial}{\partial a_{b,\tau}^t}  e^{-\Delta\tau K_{b}(x_{b,\tau}, a_{b,\tau})} (1 - G_b(\tau)) \right).
\end{equation}

\section{Gauss's law susceptibility for the OSM phase}
\label{app:gauss_suscep}

Since the $\hat{Q}_i$ is a conserved quantity it's uniform susceptibility  can be defined as 
\begin{equation}
  \chi_Q = \frac{\beta}{N}  \left\langle \left( \sum_i \hat{Q}_i \right)^2 \right\rangle. 
\end{equation}
In the  above we have omitted the background contribution since  it  vanishes due  symmetry considerations.   We now assume that  energetics of the Gauss  sectors  follow  the form  put forward in Eq.~\ref{eq:HQ_Higgs}. Thereby 
\begin{equation}
  \chi_Q  = \frac{\sum_{Q_{\text{tot}}} N(Q_{\text{tot}}) Z(Q_{\text{tot}}) e^{- \beta \lambda_Q Q_{\text{tot}}^2/N} \beta Q_{\text{tot}}^2/N}{\sum_{Q_{\text{tot}}} N(Q_{\text{tot}}) Z(Q_{\text{tot}}) e^{- \beta \lambda_Q Q_{\text{tot}}^2/N}}
\end{equation}
In the above, $N(Q_{\text{tot}}) =  \sum_{\{Q_i\}} \delta\left(Q_{\text{tot}} - \sum_i Q_i\right)$ and $Z(Q_{\text{tot}})$ is the partition function 
the partition function restricted to the sector with total  Gauss law $Q_{\text{tot}}$.  While $Q_{\text{tot}}$ takes integer values, $\phi = Q_{\text{tot}}/\sqrt{N}$ can be treated as a continuous variable in the thermodynamic limit  such that: 
\begin{equation}
  \chi_Q = \frac{\int d\phi \, \tilde{N}(\phi) \tilde{Z}(\phi ) e^{- \beta \lambda_Q \phi^2} \beta \phi^2}{\int d\phi \, \tilde{N}(\phi) \tilde{Z}(\phi) e^{- \beta \lambda_Q \phi^2}}.
\end{equation}
In a phase, the partition function is an analytical function such that we can expand it around $\phi=0$ as appropriate  for the low temperature limit.    In zeroth order, $\tilde{N}(\phi) \tilde{Z}(\phi )  \simeq \tilde{N}(0) \tilde{Z}(0)$ we obtain  a   temperature-independent result.  This is consistent with the numerical observations.

\section{Imposing Gauss's law}
\label{app:imposing_gauss_law}

We introduce an energy penalty for local violations of Gauss's law:
\begin{equation}
  \hat{H}_\lambda = \hat{H} + \hat{H}_Q, \quad \hat{H}_Q = \lambda \sum_i \hat{Q}_i^2,
\end{equation}
where \(\hat{H}\) is the Hamiltonian in Eq.~\eqref{eq:hamiltonian} and \(\lambda \geq 0\). Because the Gauss operators are conserved \([\hat{Q}_i, \hat{H}] = 0\), the eigenstates of the Hamiltonian can be chosen to have definite local Gauss charges, Eq.~\eqref{eq:eigenstates_gauss_sectors}. The \(\hat{H}_Q\) term shifts the energy according to
\begin{equation}
  E_{\{Q_i\}}(\lambda) = E_{\{Q_i\}}(0) + \lambda \sum_i Q_i^2.
\end{equation}
So the eigenstates within each sector are therefore unchanged; \(\lambda\) only changes the relative energies of the different sectors. We can also see this from the partition function:
\begin{equation}
  Z_\lambda = \sum_{\{Q_i\}} e^{-\beta\lambda \sum_i Q_i^2} Z_{\{Q_i\}}, \quad Z_{\{Q_i\}} = \text{Tr}_{\{Q_i\}} (e^{-\beta \hat{H}})  
\end{equation}
A unit local violation, \(Q_i = \pm 1\), is suppressed by a Boltzmann factor \(e^{- \beta \lambda}\). At finite \(\lambda\), non-physical sectors remain present but are exponentially suppressed. In the limit \(\lambda\to\infty\) the constraint is imposed exactly:
\begin{equation}
  \lim_{\lambda\to\infty} Z_\lambda = Z_{\{Q_i=0\}}
\end{equation}
The \(\lambda \to \infty\) limit is therefore equivalent to an exact projection onto the physical Hilbert space. Alternatively, setting \(\beta \to \infty\) exactly imposes the constraint.

To take the limit \(\lambda \to \infty\), we consider a Hamiltonian term after the Trotter decomposition and perform a continuous Hubbard-Stratonovich transformation:
\begin{equation}
  e^{-\Delta\tau \lambda \hat{Q}_i^2} = \frac{1}{\sqrt{2\pi \lambda \Delta\tau}} \int_{-\infty}^{\infty} da^0_{i,\tau} e^{- (a^0_{i,\tau})^2 / (2 \lambda \Delta\tau)} e^{i a^0_{i,\tau} \hat{Q}_i}.
\end{equation}
Then we can subdivide the integration interval in smaller intervals between \(2\pi m\) and \(2\pi (m+1)\), where \(m \in \mathbb{Z}\). Redefining \(a^0_{i,\tau} \to a^0_{i,\tau} + 2\pi m\), we have 
\begin{equation}
  \frac{1}{\sqrt{2\pi \lambda \Delta\tau}} \int_{0}^{2\pi} da^0_{i,\tau} \sum_{m \in \mathbb{Z}}  e^{- (a^0_{i,\tau} + 2\pi m)^2 / (2 \lambda \Delta\tau)} e^{i a^0_{i,\tau} \hat{Q}_i} e^{i 2\pi m \hat{Q}_i}.
\end{equation}
As the eigenvalues of the Gauss's operator are integers, \(e^{i 2\pi m \hat{Q}_i} = 1\). Using the Poisson summation formula, the summation of \(m\) becomes 
\begin{equation}
  \frac{1}{\sqrt{2\pi \lambda \Delta\tau}} \sum_{m} e^{- (a^0_{i,\tau} + 2\pi m)^2 / (2 \lambda \Delta\tau)} = \frac{1}{2\pi} \sum_n e^{i n a^0_{i,\tau}} e^{-n^2 \Delta\tau \lambda /2}.
\end{equation}
Taking \(\lambda \to \infty\), we then have 
\begin{equation}
  \lim_{\lambda \to \infty} e^{-\Delta\tau \lambda \hat{Q}_i^2} = \int_{0}^{2\pi} \frac{da^0_{i,\tau}}{2\pi} e^{i a^0_{i,\tau} \hat{Q}_i} = \delta(\hat{Q}_i).
\end{equation}
Then integration over \(a^0\) exactly reproduces the constraint \(Q_i = 0\) in the path integral formulation. 

In the same way as in Appendix \ref{app:af_qmc}, we can derive the action of \(\hat{H}_{\lambda \to \infty}\). We then have:
\begin{equation}
  Z = \int \D{a^0,a,x}  e^{-S_{\text{phonon}} - S_{\text{gauge}}} \det\left(1 + B(\beta, 0)\right),
\end{equation}
where \(S_{\text{phonon}}\) is the phonon action, Eq.~\eqref{eq:phonon_action}. The gauge action \(S_{\text{gauge}}\) is now given by 
\begin{equation}
  S_{\text{gauge}} = \sum_{\tau = 1}^{L_\tau} \left[ \frac{2}{J\Delta\tau} \sum_{i,\delta} (1 - \cos(a_{i,\delta,\tau+1} - a_{i,\delta,\tau} + a^0_{i+\delta,\tau} - a^0_{i,\tau})) + \Delta\tau K \sum_{\square} \cos(\text{curl} a_{\square,\tau}) \right],
\end{equation}
and the fermionic determinant by 
\begin{equation}
  B(\tau_2, \tau_1) = \prod_{\tau=\tau_1 + \Delta\tau}^{\tau_2} \prod_{i,\delta} e^{- \Delta\tau K_{i,\delta}(x_{i,\delta,\tau}, a_{i, \delta, \tau})} \prod_{i} e^{i a^0_{i,\tau} (n_i - 1/2)}.
\end{equation}

\end{widetext}

\bibliography{joao.bib}

\end{document}